\documentclass[times,twocolumn]{aastex631}
\usepackage{CJK}
\usepackage{mathptmx}
\usepackage[T1]{fontenc}
\usepackage{ae,aecompl}
\usepackage{graphicx}	\usepackage{amsmath}	\usepackage{amssymb}	\usepackage{subfigmat}
\usepackage{enumerate}
\usepackage{multirow}
\usepackage{natbib}
\usepackage{booktabs}
\usepackage{multirow}
\usepackage{chngcntr}
\usepackage{ulem}
\usepackage{appendix}
\usepackage{threeparttable}

\newcommand{\mdotyr}{\hbox{$M_\odot$ yr$^{-1}$}}

\received{2023 June 15}
\revised{2023 August 3}
\accepted{2023 August 16} 

\shorttitle{The Accretion History of EX Lup}
\shortauthors{Wang et al.}

\begin{document}

\title{The Accretion History of EX Lup: \\
A Century of Bursts, Outbursts, and Quiescence}
\begin{CJK*}{UTF8}{gbsn}
\author[0000-0003-3015-6455]{Mu-Tian Wang (王牧天)}
\affiliation{School of Astronomy and Space Science, Nanjing University, Nanjing, 210023, People’s Republic of China.}
\affiliation{Key Laboratory of Modern Astronomy and Astrophysics, Ministry of Education, Nanjing, 210023, People’s Republic of China}
\author[0000-0002-7154-6065]{Gregory J. Herczeg (沈雷歌)}
\affiliation{Kavli Institute for Astronomy and Astrophysics, Peking University, No.5 Yiheyuan Road, Haidian District, Beijing 100871, People's Republic of China}
\affiliation{Department of Astronomy, Peking University, No.5 Yiheyuan Road, Haidian District, Beijing 100871, People's Republic of China}
\author[0000-0001-5162-1753]{Hui-Gen Liu (刘慧根)}
\affiliation{School of Astronomy and Space Science, Nanjing University, Nanjing, 210023, People’s Republic of China.}
\affiliation{Key Laboratory of Modern Astronomy and Astrophysics, Ministry of Education, Nanjing, 210023, People’s Republic of China}
\author[0000-0001-8060-1321]{Min Fang (房敏)}
\affiliation{Purple Mountain Observatory, Chinese Academy of Sciences, No.10 Yuanhua Road, Qixia District, Nanjing 210023, People's Republic of China}
\affiliation{School of Astronomy and Space Science, University of Science and Technology of China, Hefei, Anhui 230026, People's Republic of China}
\author[0000-0002-6773-459X]{Doug Johnstone}
\affiliation{NRC Herzberg Astronomy and Astrophysics, 5071 West Saanich Rd, Victoria, BC, V9E 2E7, Canada}
\affiliation{Department of Physics and Astronomy, University of Victoria, Victoria, BC, V8P 5C2, Canada}
\author{Ho-Gyu Lee}
\affiliation{ Korea Astronomy and Space Science Institute, 776, Daedeok-daero, Yuseong-gu, Daejeon, 34055, Republic of Korea}
\author[0000-0001-7796-1756]{Frederick M. Walter}
\affil{Department of Physics \& Astronomy, Stony Brook University,
             Stony Brook NY 11794-3800, USA}
\author[0000-0003-0125-8700]{Franz-Josef Hambsch}
\affiliation{Groupe Europ\'{e}en d’Observations Stellaires (GEOS), 23 Parc de Levesville, 28300 Bailleau l’Ev\^{e}que, France}
\affiliation{Bundesdeutsche Arbeitsgemeinschaft f\"{u}r Veränderliche Sterne (BAV), Munsterdamm 90, 12169 Berlin, Germany}
\affiliation{Vereniging Voor Sterrenkunde (VVS), Oostmeers 122 C, 8000 Brugge, Belgium}
\author{Carlos Contreras Pe{\~n}a}
\affiliation{Department of Physics and Astronomy, Seoul National University, 1 Gwanak-ro, Gwanak-gu, Seoul 08826, Korea}
\author[0000-0003-3119-2087]{Jeong-Eun Lee}
\affiliation{Department of Physics and Astronomy, Seoul National University, 1 Gwanak-ro, Gwanak-gu, Seoul 08826, Korea}
\author{Mervyn Millward}
\affiliation{York Creek Observatory, George Town, Tasmania, Australia}
\author{Andrew Pearce}
\affiliation{Skynet Robotic Telescope Network, Perth Observatory, Western Australia}
\author{Berto Monard}
\affiliation{Kleinkaroo Observatory, Calitzdorp, Western Cape, South Africa}
\author{Lihang Zhou (周立杭)}
\affiliation{Department of Astronomy, Peking University, No.5 Yiheyuan Road, Haidian District, Beijing 100871, People's Republic of China}

\correspondingauthor{Mu-Tian Wang, Gregory Herczeg, Hui-Gen Liu} 
\email{mutianwang@smail.nju.edu.cn, gherczeg1@gmail.com, huigen@nju.edu.cn}

\begin{abstract}
    EX Lup is the archetype for the class of young stars that undergoes repeated accretion outbursts of $\sim 5$ mag at optical wavelengths and that last for months.  Despite extensive monitoring that dates back 130 years, the accretion history  of EX Lup remains mostly qualitative and has large uncertainties.
                We assess historical accretion rates of EX Lup by applying correlations between optical brightness and accretion, developed on multi-band magnitude photometry of the 
                $\sim 2$ mag optical burst in 2022.
    Two distinct classes of bursts occur: major outbursts ($\Delta V\sim5$ mag) have year-long durations, are rare, reach accretion rates of $\dot{M}_{\rm acc}\sim10^{-7}$ \mdotyr\ at peak, and have a total accreted mass of around 0.1 Earth masses. The characteristic bursts ($\Delta V\sim2$ mag) have durations of $\sim 2-3$ months, are more common, reach accretion rates of $\dot{M}_{\rm acc}\sim10^{-8}$ \mdotyr\ at peak, and have a total accreted mass of around $10^{-3}$ Earth masses.
The distribution of total accreted mass in the full set of bursts is poorly described by a power law, which suggests different driving causes behind the major outburst and characteristic bursts.
    The total mass accreted during two classes of bursts is around two times the masses accreted during quiescence.  Our analysis of the light curves reveals a  color-dependent time lag in the 2022 post-burst light curve, attributed to 
    the presence of both hot and cool spots on the stellar surface. 
\end{abstract}


\section{Introduction}
\end{CJK*}
Outbursts of accretion onto young stars play a significant role in the assembly of stellar mass \citep{fischer22} and the chemistry of any envelope and the protoplanetary disk \citep[e.g.][]{lee07,jorgensen15,hsieh19}.  
Eruptive young stellar objects (YSOs) are conventionally classified by their photometric and spectroscopic characteristics into two categories, FU Ori-type and EX Lup-type objects, after their namesakes.  The FU Ori-type outbursts reach accretion rates of $10^{-6}-10^{-4}$ \mdotyr, a factor of $10^3-10^5$ higher than the quiescent accretion rates of $\sim 10^{-9}-10^{-8}$ \mdotyr, and can last for decades or even centuries, while the smaller EX Lup-type outbursts have accretion rates of $\sim 10^{-7}$ \mdotyr\ that last for months \citep[see reviews by][]{audard14,hartmann16}.  Follow-up spectroscopy is often used to classify EX Lup-type outbursts from emission lines \citep[e.g.][]{connelley18}, though modern searches reveal that many accretion outbursts have characteristics intermediate between the EX Lup and FU Ori classes \citep[e.g][]{contreras17,guo21}.

We focus on EX Lup, the namesake of a class of outbursts and a member of the Lupus star-forming region \citep[e.g.][]{alcala17} that is known for repeated bursts.  Monitoring since 1893 revealed large outbursts in 1945, 1955, and 2008 \citep[e.g.][]{mclaughlin46,Herbig92,Aspin2010}, marking bookends for the studies of EX Lup by George Herbig \citep{herbig1950,herbig2008}.  Interspersed between these outbursts were many smaller ($\Delta V\sim 2-3$ mag) bursts that lasted for at most a few months 
\citep{Herbig2001,Herbig2007}.
 The accretion rates  have been estimated to range from $10^{-8}-10^{-10}$ \mdotyr\  in quiescence \cite[e.g.][]{Sipos2009,Sicilia-Aguilar2015,alcala17} and from $10^{-7}-10^{-6}$ \mdotyr\ for the large outbursts (e.g., \citealt{Aspin2010,Juhasz2012}, see also \citet{giannini22} for other sources), but with differences between studies of 1--2 orders of magnitude.
 
 Measurements of the accretion rates during both quiescence and outburst provide a quantified mass measurement that can be used to constrain the instability physics. For EX Lup, the outbursts are likely caused by instabilities in the innermost disk or in the magnetic star-disk connection \citep[e.g.][]{dangelo10,armitage16}.  
 The largest outbursts can also have a significant impact on the disk structure and chemistry. 
 The 2008 outburst led to the depletion of gas in the inner disk \citep{Banzatti2015}, moved snowlines to larger radii \citep{White2020}, and created comet-forming materials  \citep{Abraham2009}.  Organic molecules reformed after destruction during the outburst, while crystalline grains that formed during the burst were transported outward \citep{kospal23}. 

The most recent burst of EX Lup, in 2022 and initially identified by \citet{zhou2022} and \citet{kospal22}, was intensely monitored with multi-band photometry, including by astronomers who contributed to the database of the American Association of Variable Star Observers (AAVSO). The exquisite light curve provides an opportunity to study the changes in brightness and mass accretion rate.  Correlations between brightness and accretion rate during the 2022 burst aid further 
estimates of historical accretion rates from the long-term photometric light curve. Our results are mostly similar to the contemporaneous and independent analysis of the 2022 burst and historical light curve by Cruz S\'aenz de Miera et al. 2023.

In this paper, we re-evaluate the accretion history of EX Lup based on correlations developed from the burst in 2022.  As the archetype of its class and with historical monitoring that dates back to 1893, the accretion history can be reconstructed for over a century.  With the improved knowledge about the variability link between mass accretion rate and photometry, we return to the century-long photometry data of EX Lup to measure the mass accreted in bursts and compare that to the mass accreted during quiescence.  In Section \ref{sec:photometry_history} we describe the historical and modern photometry and spectroscopy of EX Lup, and review the stellar properties of EX Lup in Section \ref{sec:stellar_properties}. In Section \ref{sec:photometric_overview} we provide a photometric analysis of EX Lup's historical and modern light curves. In Section \ref{sec:excess_emission_acc_rate} we describe our method to translate multi-wavelength photometry to mass accretion rates. In Section \ref{sec:historical_bursts} we estimate the accreted masses during burst and in quiescent.

\section{Historical and Recent Light Curves and Spectra \label{sec:photometry_history}}

\begin{figure*}
    \centering
    \includegraphics[width=1\textwidth]{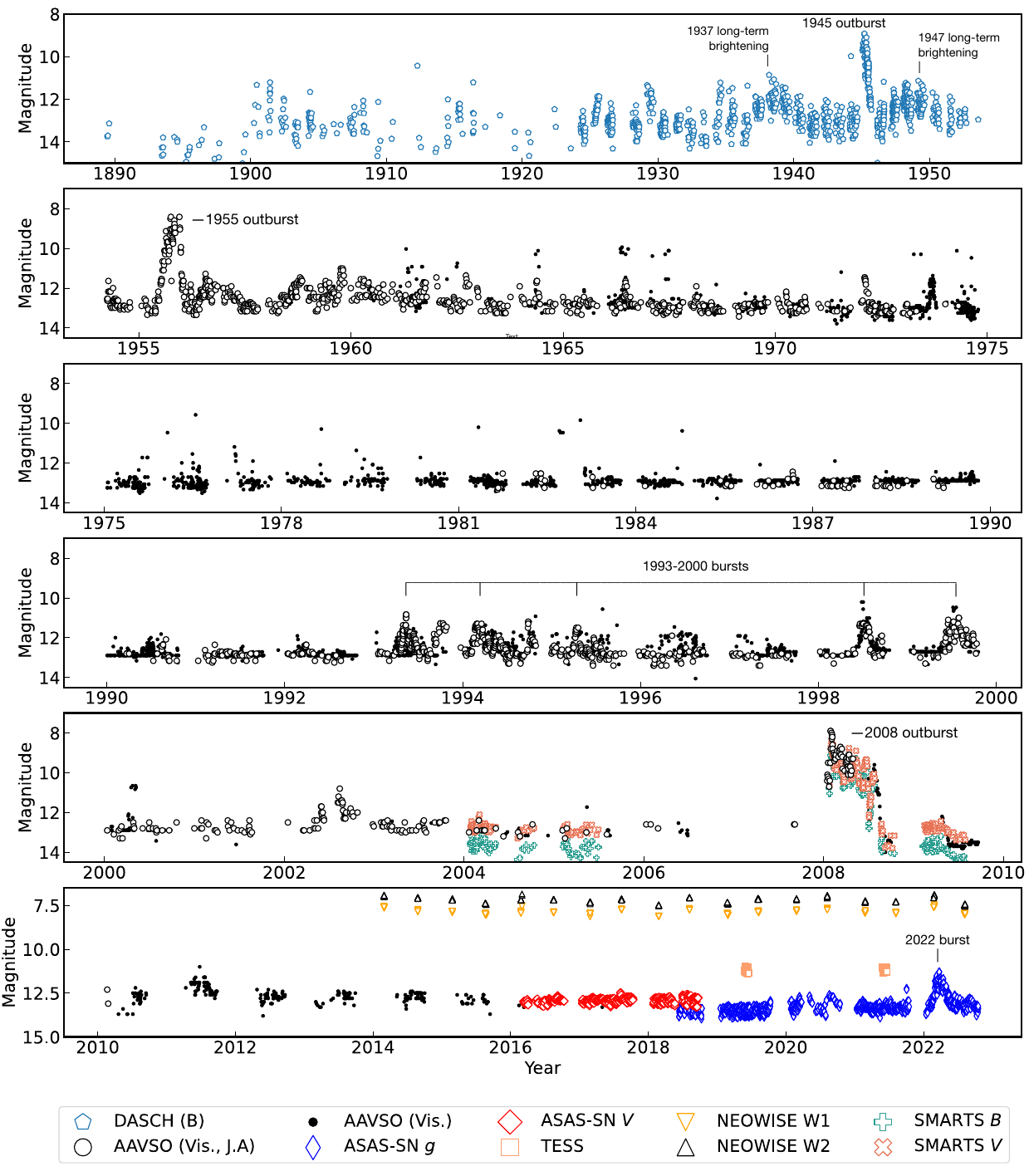}
    \caption{The full light curve collections of EX Lup. The photometry data covered in this paper come from DASCH (blue petagons), AAVSO (black dots), ASAS-SN (red and blue diamonds), NEOWISE (black and orange triangles), TESS (orange squares), and SMARTS/1.3-m telescope (red and green crosses). The timestamps of several outbursts and bursts from EX Lup are also indicated in the figure.}
    \label{fig:full_lightcurve}
\end{figure*}

This section summarizes the photometric and spectroscopic data we obtained for analysis in this paper. The photometric data is shown in Figure \ref{fig:full_lightcurve}.

\subsection{DASCH light curves}
We obtain the EX Lup $B$-band light curves available between 1893 and 1954 from the DASCH database.  The Digital Access to a Sky Century@Harvard \citep[DASCH, ][]{Laycock10,Tang13} is a project to digitize historical data from a collection of photographic plates hosted by the Harvard College Observatory between 1885 to 1992. The plate-derived photometry was further calibrated by AAVSO Photometric All-Sky Survey \citep[APASS][]{zacharias13} and is converted into Johnson B magnitudes. 
\footnote{http://dasch.rc.fas.harvard.edu/lightcurve.php}

\subsection{AAVSO Light Curves}

American Association of Variable Star Observers (AAVSO) database has a wealth of photometry of EX Lup from 1955 to 2022. We obtained the full collection of EX Lup photometry from the AAVSO website\footnote{https://www.aavso.org/} on Nov.~10 2022. The majority of AAVSO photometry is in the format of Visual magnitude before 2008, sourced from the Royal Astronomical Society of New Zealand (RASNZ) and in the non-modern format of the visible band, as observed by different observers, most especially Albert Jones in the Visual band before 1970 and between 1987-2007.

\cite{Herbig92} transformed the 1955-1956 Visual photometry from Jones to the Johnson V band by
\begin{equation}
\label{eq:vis2v}
    V = 0.895\times Vis. + 0.63.
\end{equation}
We extrapolate this relation to photometry from Jones between 1955-1993. After 1994, we infer that Jones reported photometry in the standard Johnson V magnitude, because the minimum visual magnitude changed from $\sim$14 to $\sim$13, close to the minimum V band magnitude (13.25) mag observed by ASAS-SN.  

The recent 2022 burst was also covered with frequent full-length multi-color $UBV(RI)_c$ photometric monitoring. Based on the record from AAVSO, the campaign starts on 2022-03-13 and ends on 2022-10-17.  The U-band photometry, particularly powerful for accretion rate measurements  \citep{gullbring98}, was reported along with $BV(RI)_c$ by Franz-Josef Hambsch; 14 other observers also contributed to this light curve. Thanks to contributions from all the observers, the seven-month multi-color photometry monitoring is kept on a daily cadence, with very few one-day gaps.

\subsection{ASAS-SN Light Curves}

The All-Sky Automated Survey for Supernovae (ASAS-SN) catalog  \citep{Shappee2014,Kochanek2017} provides $V$-band photometry from 2016-3-10 to 2018-9-17 and $g$ band photometry from 2018-6-13 to present, downloaded on 2023-07-17 from the ASAS-SN website\footnote{https://asas-sn.osu.edu/}.  On a typical night, ASAS-SN will observe EX Lup three times with 90-second exposures, but sometimes there are cases where only one or two exposures are available because of the weather. These differences have a negligible effect on the photometric analysis presented in Section \ref{sec: asassn_tess_spot}.   

The ASAS-SN camera has a field of view of $4.5\times4.5^{\circ}$ and a pixel size of 8$''$.  The contamination by other stars is negligible within the large aperture. One source separated by $8\farcs7$ has Gaia $G=16.4$ mag, roughly 4 mag fainter than EX Lup, while all others are fainter than 18 mag \citep{gaiaedr3}.

\subsection{TESS Photometry}
EX Lup was observed by TESS in Sector 12 and 39 with a 20-second cadence. We downloaded the \texttt{SPOC}-processed TESS light curves of EX Lup in these two sectors, and convert the TESS flux in units of electron per second $f_e$ to TESS magnitude $T = -2.5\log(f_e) + 20.44$ from the formula in TESS Instrument Handbook.  Although the  pixel scale of TESS is 21$^{\prime\prime}$, EX Lup dominates the total fluxes within the one-pixel range. The TESS filter extends from $\sim 6000-10000$ \AA.

\subsection{NEOWISE Photometry}

The mid-IR photometry arises from all-sky observations of the {\it WISE} telescope. {\it WISE} surveyed the entire sky  from 2010  to 2011 in bands W1 (centered at 3.4 $\mu$m) and W2 (centered at 4.6 $\mu$m), with spatial resolutions of 6.1\arcsec~and 6.4\arcsec, respectively \citep{2010Wright, 2011Mainzer}. In 2013 September, WISE was reactivated as the NEOWISE-reactivation mission \citep[NEOWISE-R,][]{2014Mainzer}. NEOWISE-R is still operating and the latest released data set contains observations until mid-December 2022. For each visit to a particular area of the sky, {\it WISE} performs several photometric observations over a period of $\sim$few days. Each area of the sky is observed in a similar way every $\sim$ 6 months. The data presented were obtained between 2010 and 2022 and downloaded from the single-epoch catalogues stored at the NASA/IPAC Infrared Science Archive (IRSA) catalogues \citep{https://doi.org/10.26131/irsa144}, using a 3\arcsec~radius from the coordinates of the YSO.

\subsection{Optical SMARTS/Andicam Photometric Monitoring}

EX Lup was observed in Johnson $B$ and $V$ band by SMARTS/Andicam on the 1.3-m telescope \citep{SMARTS} from 2004-01-27 to 2005-06-29 in quiescence on a weekly cadence, and from 2008-01-22 to 2009-07-30 during outburst on nightly cadence. In total, SMARTS/Andicam obtained 297 $B$-band observations on 292 nights and 286 $V$-band observations on 286 nights.
 
\subsection{Optical-near-IR VLT/X-Shooter spectrum}

EX Lup was observed by VLT/X-Shooter on UTC 2010-05-04 03:44:51.779 as part of program 085.C-0764 (PI H.M. G\"unther) and was initially published by \citet{alcala17}.   The X-Shooter spectrum is flux-calibrated and covers from 3000--25000 \AA.
The synthetic magnitudes derived for the X-Shooter spectrum are Johnson U=13.927, V=12.956, B=14.087, SDSS g = 13.633 mag, 2MASS J = 9.80 mag, after applying filter curves from the Spanish Virtual Observatory \citep{rodrigo12,rodrigo20}. The estimated extinction-corrected and accretion-subtracted magnitude of the photosphere is U = 15.474, V=13.055, B=14.535, SDSS g = 13.892 mag, 2MASS J = 9.89 mag with typical uncertainties of $\sim 0.05$ mag.  At $J$-band, this indicates a veiling of $\sim 8$\% during quiescence, similar to $J$-band veiling measurements for other accreting low-mass stars \citep{fischer11,sousa23}.

\subsection{Optical Keck/HIRES spectrum}

We obtained a flux-calibrated high-resolution spectrum of EX Lup with Keck I/HIRES on MJD 54609.416 (2008 May 23), during the large burst in 2008, originally published in \citet{Aspin2010}.  Fluxes were calibrated using two spectra of the spectrophotometric standard EG 274 \citep{hamuy92}, both obtained within $1$ hr of the EX Lup spectra and bracketing the airmass of the EX Lup observation.

\subsection{Near-IR Gemini South/IGRINS spectrum}

We obtained a high-resolution ($R\sim 45,000$) spectrum of EX Lup from 1.4--2.5 $\mu$m 
using Immersion Grating Infrared Spectrograph \citep[IGRINS, ][]{park14} at Gemini South 
on 2022-03-14 (MJD 59652.36), obtained in GS-2022A-Q-109 (PI H-G Lee) through the 
Korean partnership in Gemini Observatory.  The observations were obtained in  
2$\times$ABBA pattern with 131-second integrations at each step.  
The A0 star HIP78445 was observed as a standard star to correct for 
the telluric absorption.  The data were reduced using the standard IGRINS pipeline \citep{lee17} with a modified version of optimal extraction \citep{horne86}.

The flux in the EX Lup spectrum was calibrated from the relative count rate in the EX Lup and telluric standard star.  For both EX Lup and HIP 78445, the slit viewer indicates that the stars were well centered.  We also corrected for differences in slot losses by measuring the spatial profile in the cross-dispersion direction and assuming a Gaussian point-spread function.  From this process, we measure $K=8.07$ with an uncertainty of $\sim 0.1$ mag.

\begin{figure}
    \centering
    \includegraphics[width=0.5\textwidth]{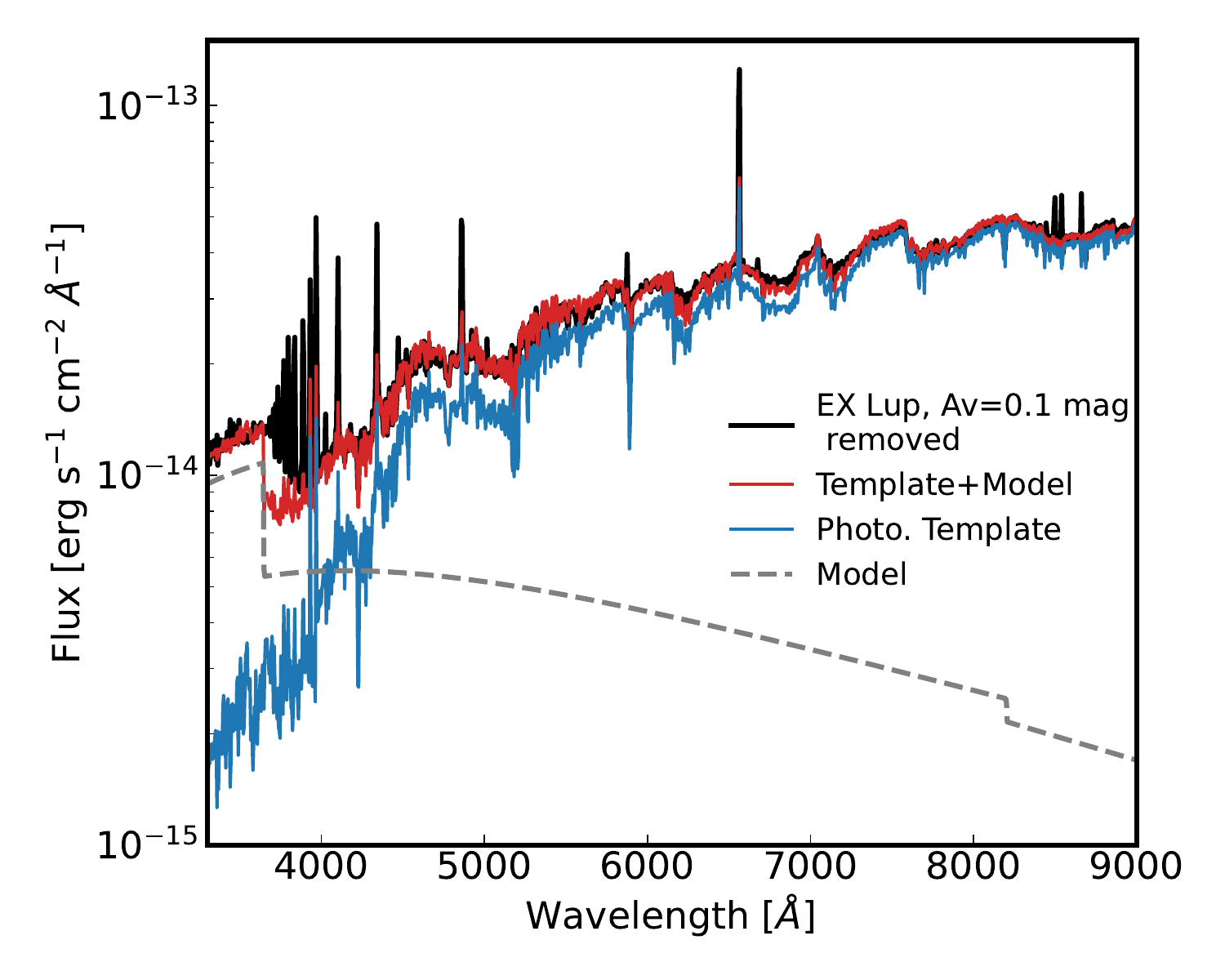}
    \caption{The best-fit (red) of EX Lup X-Shooter spectrum (black) taken on May 4, 2010. The spectrum is dereddened with $A_v=$ 0.1 mag, fitted with a photosphere template (blue) and a slab model (grey).}
    \label{fig:exlup_spec}
\end{figure}

\begin{figure}
    \centering
    \includegraphics[trim={3cm 13cm 1.8cm 3cm},width=0.47\textwidth]{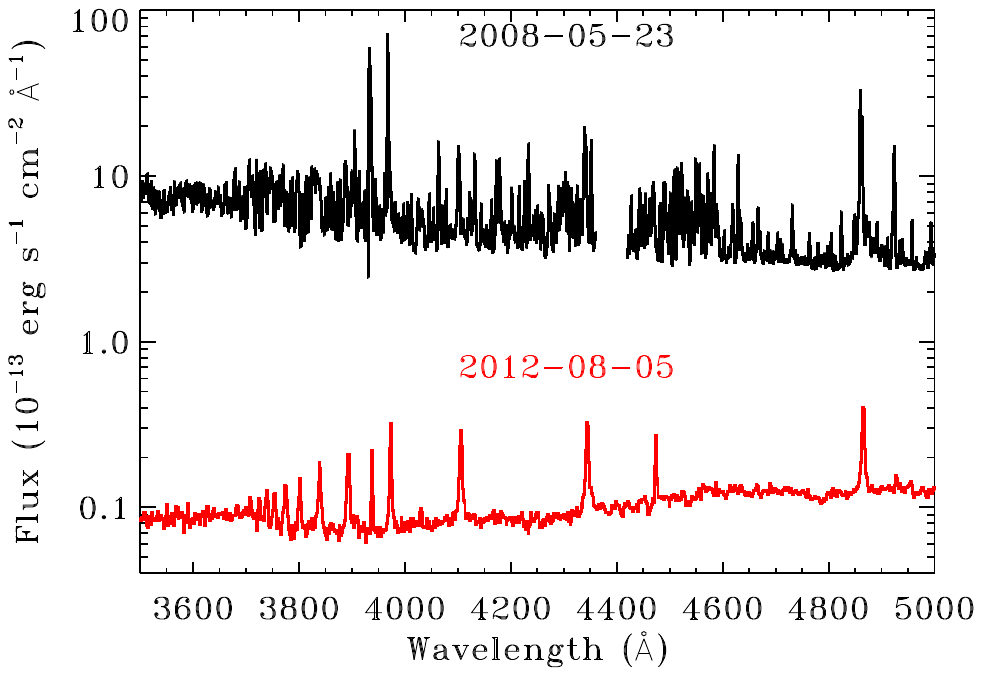}
    \includegraphics[trim={3cm 13cm 1.8cm 3cm},width=0.47\textwidth]{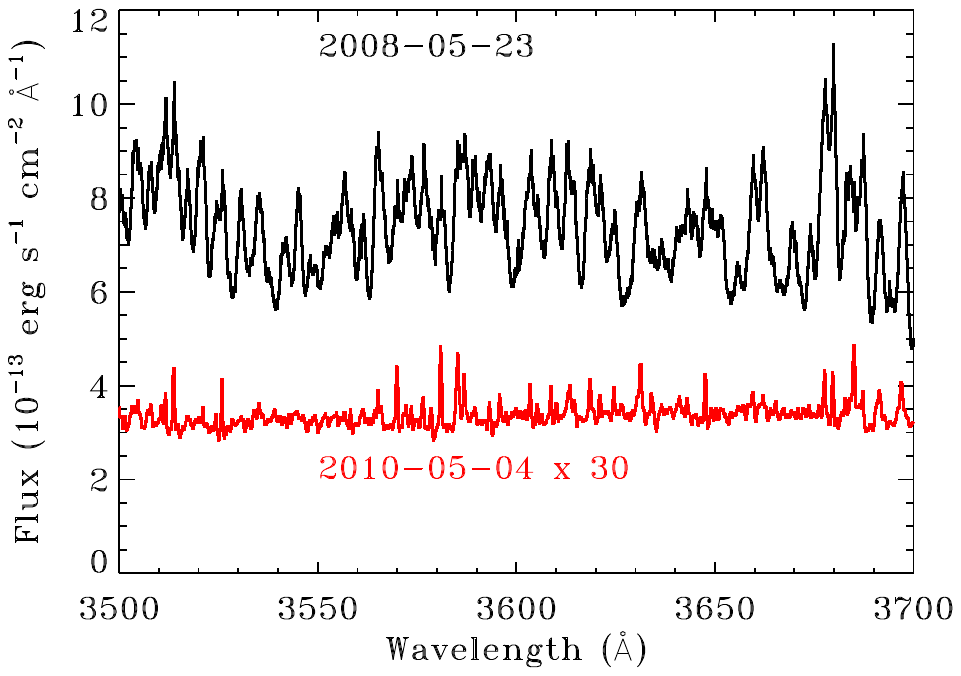}
    \caption{The Balmer Jump (top) and U-band spectra (bottom) for EX Lup in quiescence (red) and outburst (black).  The accretion spectrum of EX Lup during outburst has a bluer slope and maintains a Balmer Jump.  The U-band spectrum during the outburst is highly structured, indicating that the accretion flow has an optically thick photosphere.  In quiescence, the U-band spectrum of EX Lup (and most other accreting young stars) is mostly featureless.}
    \label{fig:exlup_2008jump}
\end{figure}

 \section{Properties of EX Lup\label{sec:stellar_properties}}

In this section, we re-evaluate the properties of EX Lup, including revising the extinction from $A_V=1.1$ mag in \cite{alcala17} to $0.1$ mag based on an analysis of the same X-Shooter spectrum.  A more complete analysis of the X-Shooter spectrum and our revised parameters is presented in Appendix A.

The optical spectrum of EX Lup in quiescence is the combination of a cool stellar photosphere, including TiO bands in absorption, and excess accretion continuum and line emission.  To determine the accretion components in the spectrum, we test several non-accreting, low-mass YSO spectra to serve as photospheric templates of EX Lup \citep{manara13}.  We find the most appropriate photospheric template from \cite{manara13} and extinction that yields an accretion continuum spectrum that is most free of photospheric features (see Appendix for details).

Our final measurements are a spectral type of M0.5 (corresponding to $3810$ K using the conversion of \citealt{herczeg14}) and extinction $A_V=0.1$ mag, determined by extinction curve of $R_V=3.1$ \citep{ccm89}. This fit yields a residual (accretion) spectrum with no evident molecular band absorption features but with fluctuations that are larger than seen for many other accreting young stars.  The dereddened EX Lup spectrum, the photospheric spectrum template (TWA 14), and the subtracted accretion spectrum are displayed in Figure \ref{fig:exlup_spec}.

In contrast to our spectral fits, \cite{alcala17} measured M0 and an extinction $A_V=1.1$ mag, estimated using a K7 template.  The lower extinction from our work is consistent with the expectation of low extinction from the non-detection of DIBs \citep[see correlations in][]{carvalho22} in both in-burst and quiescent spectra.  The spectral type is roughly consistent with the effective temperature measurement of $3750$~K by \citet{Sipos2009}.

When combined with the distance of $154.72\pm0.36$ pc \citep{2022arXiv220800211G}, the extinction- and veiling-corrected photospheric $J$-band magnitude of 9.89,
and bolometric correction from \citet{pecaut13} for an M0.5 star, we obtain a luminosity of $0.43$ L$_\odot$ and radius $1.50$ R$_\odot$.   In the \citet{somers20} pre-main sequence models, these parameters correspond to a mass of 0.83 M$_\odot$ and age of 4.6 Myr for 50\% spot coverage, which is adopted here. The 0\% spot models correspond to 0.55 M$_\odot$ and 2.3 Myr.

 \subsection{The spectroscopic accretion rate during quiescence \label{sec: optical_spec_quies}}
 
From the X-Shooter spectrum, the accretion luminosity is $0.0256$ L$_\odot$, as measured from the excess continuum from 3500--5000 \AA.  The bolometric correction is calculated from a plane-parallel slab model of \citet{valenti93}, as implemented by \citet{herczeg08}.  This accretion luminosity leads to an accretion rate of $1.77\times10^{-9}$ M$_\odot$ yr$^{-1}$.

The accretion luminosity measured here is 8 times smaller than the 0.2 L$_\odot$ and the accretion rate is 20 times smaller than the $3.9\times10^{-8}$ from \citet{alcala17}, using the exact same spectrum. 
However, most of this difference is attributed to the higher extinction of $A_V=1.1$ mag in \citet{alcala17}, which changes the accretion luminosity by a factor of $\sim 4.3$ and the radius estimate from red-optical emission by $\sim 1.5$.  The larger pre-Gaia distance of 200 pc, adopted by \citet{alcala17}, accounts for a factor of 1.7 in accretion luminosity and 1.3 in stellar radius.

\subsection{The accretion rate during the 2008 outburst}

The optical spectrum of EX Lup in late May 2008, near the end of the large outburst (Figure~\ref{fig:exlup_2008jump}), consists of a continuum with a forest of emission lines.  The emission line spectrum, also reported by \citet{Aspin2010} and \citet{Sicilia-Aguilar2012}, is similar to the optical spectra of other heavily veiled, high accretion rate young stars \citep[e.g.][]{hamann92,gahm08}.  During quiescence, these lines are weak but detectable, after subtracting photospheric templates \citep{Sicilia-Aguilar2015}.
The accretion continuum during the burst spectrum of 2008 is much bluer than the accretion continuum during quiescence, suggestive of the outburst producing a hotter temperature in the accretion shock region where the optical emission is produced.

 The continuum emission at $<3700$ \AA\ is brighter than that at longer wavelengths, consistent with expectations for the Balmer jump that characterizes magnetospheric accretion \citep[e.g.][]{calvet98}.   When modeled as H continuum emission, the accretion luminosity is $2.00$ L$_\odot$, leading to a mass accretion rate of $1.5\times10^{-7}$ M$_\odot$ yr$^{-1}$.   The stellar photosphere is no longer detectable, so all of the optical emission is now attributed to accretion.  
 
 While typical accreting young stars, including EX Lup in quiescence, have Balmer continuum spectra that are mostly featureless, the 2008 burst spectrum features a complex array of absorption features short of the Balmer jump.   The accretion flow seems to be sufficiently optically thick for photosphere-like absorption lines to appear.
 An analysis of the detailed spectral features is beyond the scope of this paper.

\subsection{Spectroscopic classification of bursts and outbursts}

Spectroscopy is historically used to discriminate between types of bursts.  Bona fide EX Lup-type outbursts show strong CO overtone emission, indicating a warm disk surface, along with strong Br$\gamma$ emission produced by magnetospheric accretion \citep[see, e.g.,][]{lorenzetti12,connelley18,contreras23,ghosh23}.  Weaker accretion bursts should produce Br$\gamma$ emission but are not expected to show CO emission; in stronger bursts the magnetospheric accretion paradigm may break down \citep[e.g.][]{hartmann96}. 

The major outburst of EX Lup in 2008, the archetype of EX Lup-type outbursts, featured strong CO overtone emission and Br$\gamma$ emission \citep[e.g.][]{Aspin2010,kospal11}.  The strength of the CO emission was much weaker at the end of the burst than at the beginning (\citealt{Aspin2010}, see also changes in fundamental emission in \citealt{goto11}), either because of a decrease in the gas in the inner disk \citep{Banzatti2015} or a decrease in heating of the disk surface.  
 
The 2022 burst shows CO in absorption, consistent with the classification as a characteristic burst rather than a major EX Lup-type outburst.  While the depth of CO bands is shallower during the 2022 burst than during quiescence in 2010, this change is explained by a veiling increasing by 0.65 at 2.3~$\mu$m.  This increase in veiling is more than expected from the synthetic K-band brightness (8.46 in 2010 compared with 8.31 during the 2022 burst), even accounting for stronger dust emission and 10\% fainter photospheric emission at 2.3~$\mu$m than at 2.15 $\mu$m.  Although no CO emission is detected, we cannot rule out the presence of weak CO emission on top of the photospheric absorption.

\section{Analyzing Photometry from Old and New Bursts\label{sec:photometric_overview}}

In this section, we begin by describing the historical brightness of EX Lup, following previous descriptions \citep[e.g.][]{mclaughlin46,Herbig92,Herbig2007}.   We then update this description for modern monitoring campaigns by ASAS-SN and by other astronomers, followed by the detailed analysis of a burst in 2022.

\subsection{Overview of the Historical Lightcurve of EX Lup}

The photometric monitoring of EX Lup began with plate photometry by Harvard Observatory in 1890, which is displayed in the top panel of Figure \ref{fig:full_lightcurve}. From 1890-1920, EX Lup had a minimum brightness of 14 mag in the B-band, 
accompanied by irregular fluctuations between 14-11 mag \citep{mclaughlin46}. As the observational cadence increased between 1920 and 1955, the photometric behavior of EX Lup became more apparent. A long-term brightening event ($\Delta B\sim 2$ mag) began in 1937 and lasted for 3-4 years, followed by a year-long outburst with $\Delta B \sim 4$ in 1945. This photometric variability pattern repeated again between 1947-1956, with similar magnitude change and durations of long-term brightening and outburst compared to the previous cycle, including another large outburst in 1955 \citep{Herbig92}. 

In the decades after 1955, EX Lup was more frequently observed by the AAVSO.  From 1955--1975, EX Lup had a base level of 
$V=13.5$ mag, with sporadic bursts with $\Delta V\sim 2$ mag and durations of less than a year \citep{bateson91}. The bursts were then rare from 1975-1993 as seen in third panel of Figure \ref{fig:full_lightcurve}. Activity resumed with seven bursts between 1993--2002, each 
with a duration and photometric amplitude similar to the bursts in the 1960s and intervals of around 1-2 years  \citep{Lehmann1995, Herbig2001, Herbig2007}. Another large outburst occurred in 2008, with a 300-day duration and 5 magnitudes change in the V-band \citep{Aspin2010,Juhasz2012}. 

The historical photometry from DASCH and AAVSO reveals a complex ensemble of photometric variability from EX Lup. 
When in quiescence, EX Lup has a night-to-night range in the variability of $<0.6$ mag in $g$-band, typical of other accreting young stars \citep[e.g.][]{venuti21,wangxl23}.
The bursts with amplitudes of $\sim 2$ magnitudes and durations of a few months are more common than the outbursts with amplitudes of 4-5 magnitude and durations of $\sim 6$ months.

\subsection{Photometric Variability Analysis From Recent Light Curves \label{sec: asassn_tess_spot}}

\begin{figure*}[t]
    \centering
    \includegraphics[width=0.9\textwidth]{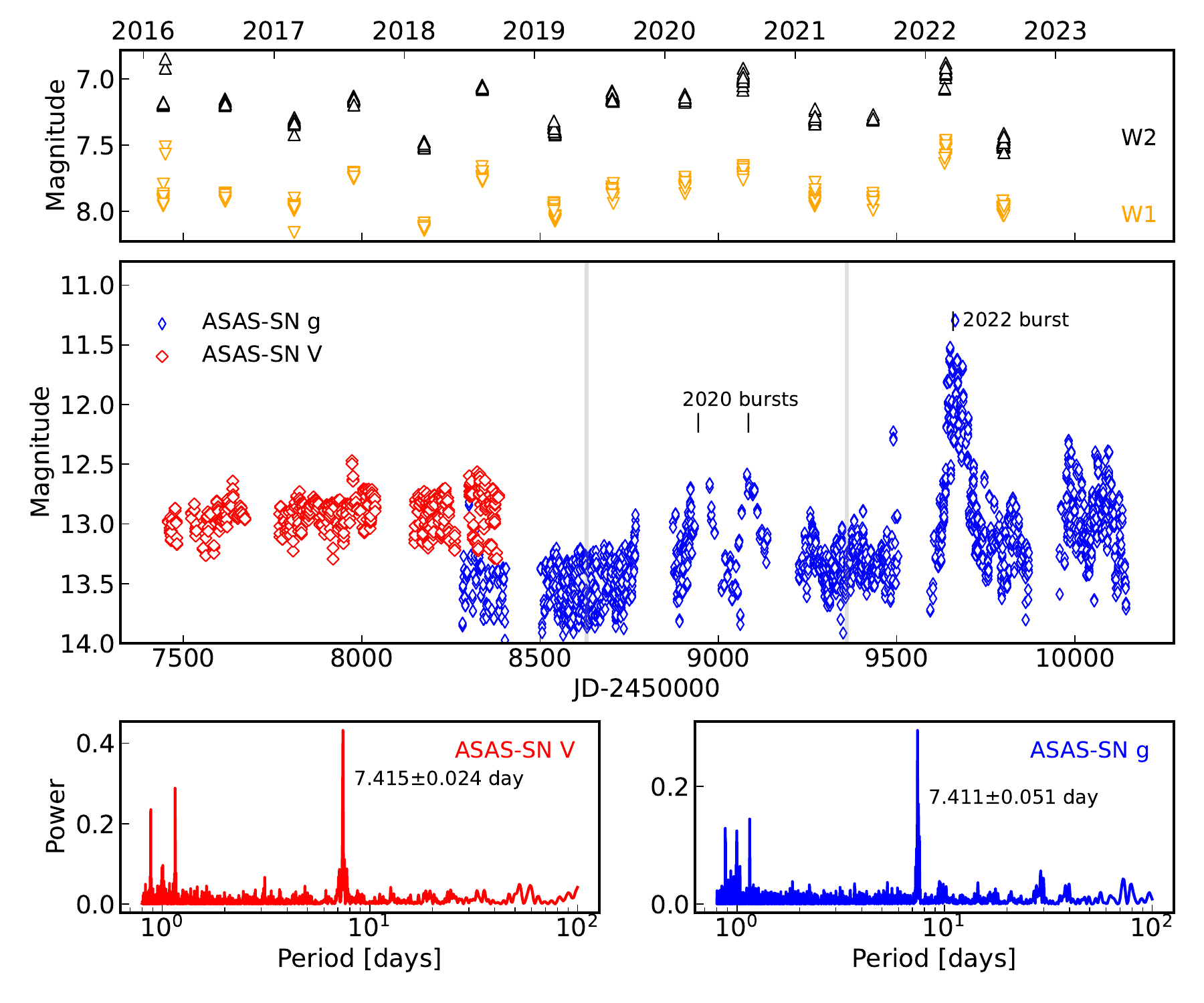}
    \caption{Top: NEOWISE $W_1$ (orange) and $W_2$ (black) photometry. Middle: The ASAS-SN $g$- (red) and $V$-band (blue) light curves of EX Lup between 2015 and 2022, shown in blue and red dots, respectively. The bursts in 2020 and 2022 are indicated with black sticks. The two gray vertical lines indicate the epochs when two sectors of TESS observations are taken. Bottom: the Lomb-Scargle periodograms for $V$-band (left) and $g$-band (right) light curves. The $g$-band light curves in 2020 and 2022 are excluded from periodogram analysis due to the presence of bursts.}
    \label{fig:assasn_wise}
\end{figure*}

\begin{figure*}
    \centering
    \includegraphics[width=0.9\textwidth]{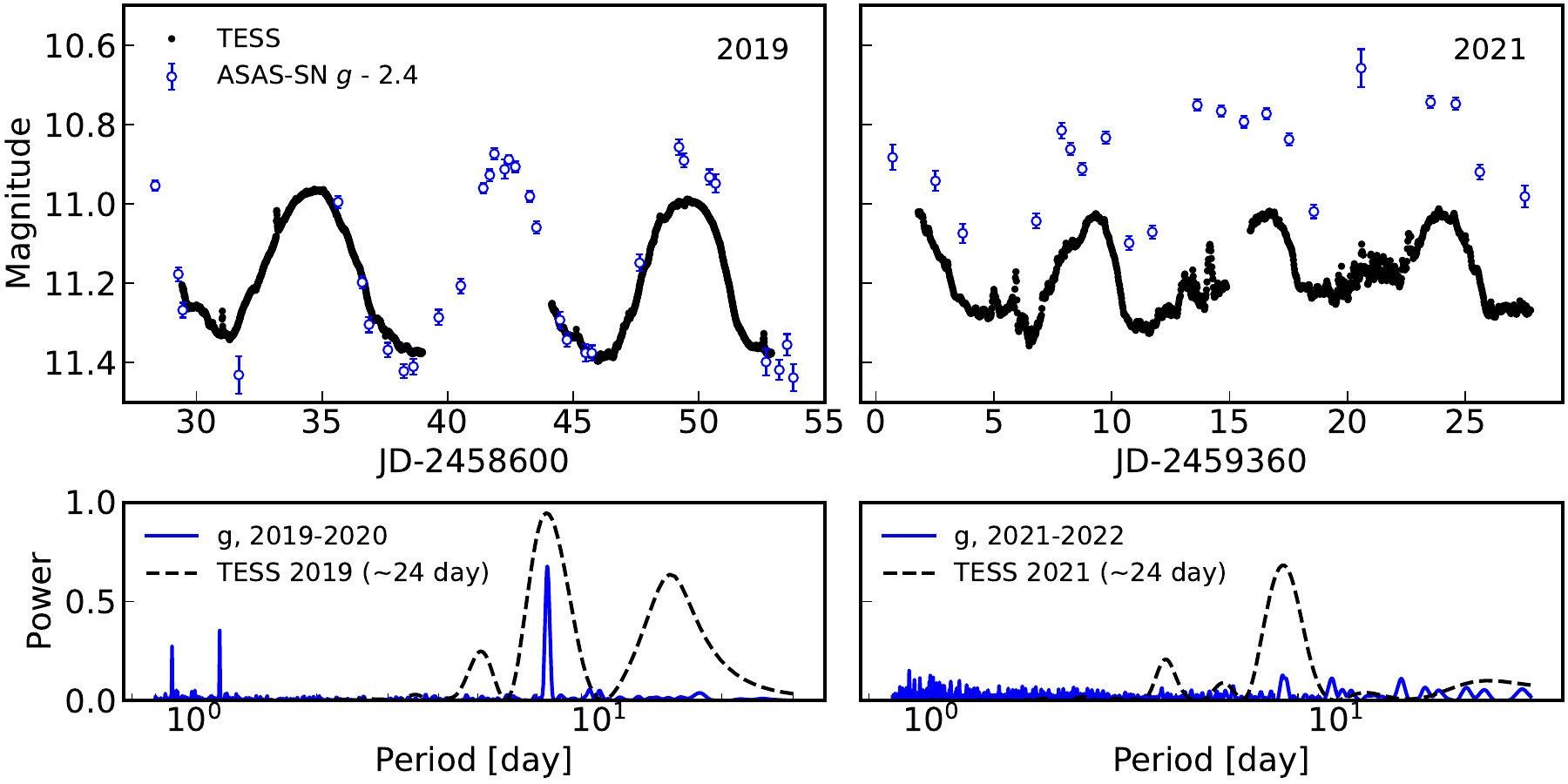}
    \caption{Top: the TESS light curves in 2019 and 2021. Contemporaneous ASAS-SN $g$-band light curves are also plotted with offset. For the clarity of display, $g$-band photometry taken within one day is binned as one, and TESS light curves are binned to a 10-minute cadence. Bottom: the Lomb-Scargle Periodogram for TESS during two sectors (each $\sim$ 24 days) and ASAS-SN $g$-band light curves in 2019-2020 and 2021-2022. }
    \label{fig:tess}
\end{figure*}

The ASAS-SN light curve taken between 2015 and 2023 traces almost a decade of the behavior of EX Lup with a near-nightly cadence. 
The $V$-band light curve between 2015 and 2018 shows a slow brightening of 0.1 mag in mean brightness and no bursts. 
Two consecutive bursts are also observed in 2020, though with a sparse cadence. The larger 2022 burst is well-monitored in $g$-band.  The $g$-band light curve during quiescent periods of 2021 has a median value that is 0.2 magnitude brighter compared to the light curve in 2019. 

TESS photometry allows us to compare the light curves before and after the 2020 bursts in greater detail (Figure \ref{fig:tess}). The TESS light curves in 2019 show periodic modulations with a peak-to-peak amplitude of 0.4 mag. The $g$-band light curves also follow a periodic modulation with an amplitude of 0.5 mags.  After the 2020 burst, the periodic modulation is still present in the TESS light curves, but additional stochastic bursts emerge in the brightness minima.  The
$g$-band photometry is brighter in 2021 than in the 2019 TESS visit, indicating a higher accretion rate. As a result, the 2021 TESS lightcurve has a higher minimum brightness, with accretion filling in the minimum brightness from the photosphere, and shows several small, readily detected bursts.

\begin{figure*}[!t]
    \centering
    \includegraphics[width= 0.95\textwidth]{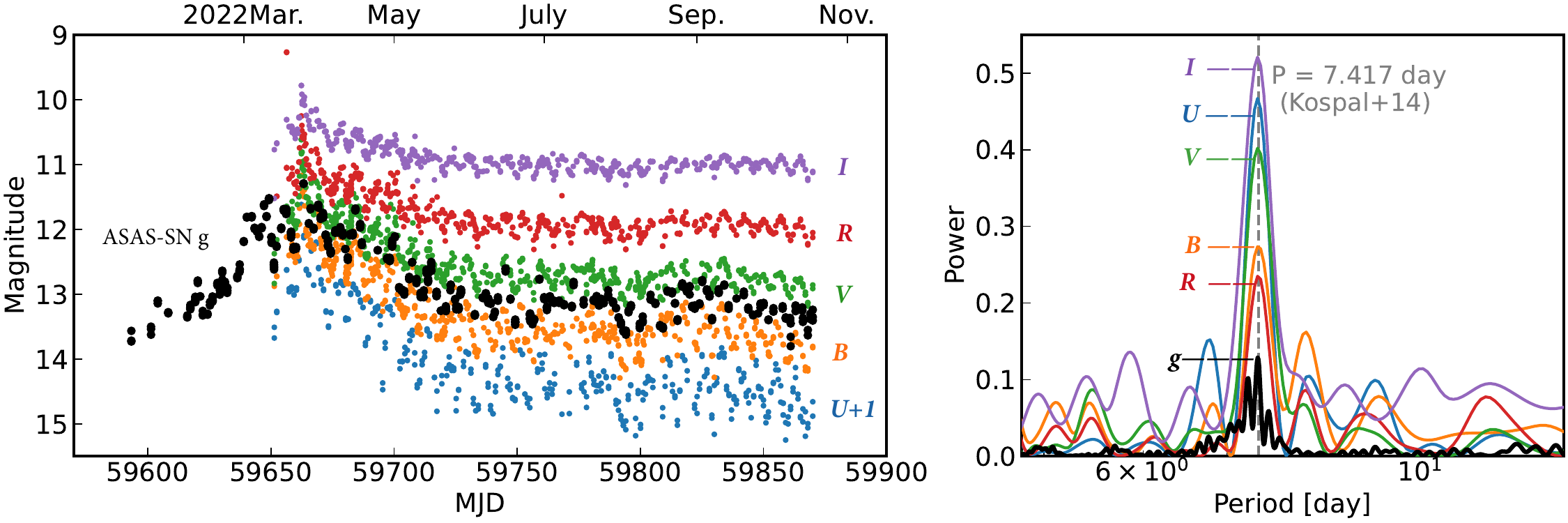}
    \caption{Left: the ASAS-SN $g$ and AAVSO multi-color light curves of 2022 burst. The light curves are offset with different values for clarity. Right: Lomb-Scargle periodogram of post-burst (after MJD=59725.0) light curves, the color scheme is the same as the left plot. The 7.417-day periodicity discovered in radial velocity measurements is also indicated \citep{Kospal2014}. }
    \label{fig:2022_burst_photometry}
\end{figure*}

The top panel of Figure~\ref{fig:assasn_wise} shows the contemporaneous NEOWISE W1 and W2 mid-infrared photometry. The mid-IR emission brightened with the 2020 and 2022 bursts, although the mid-IR emission was also brighter by 0.3 mag in epochs in 2017 and 2018 without any evident bursts in ASAS-SN light curves. We further discuss the connection between mid-IR emission and mass accretion rates in Section \ref{sec:accrate_var}.

A Lomb-Scargle Periodogram analysis of the ASAS-SN lightcurve shows a peak of 7.415$\pm$0.024 days for the $V$-band and 7.411$\pm$0.051 days for the $g$-band (bottom panels of Fig.~\ref{fig:tess}). We found no evidence of a change in periodicity in the light curves taken each year, though the power of the peak periodicity depends on the level of burst activities in each year. For 2019, the Lomb-Scargle periodogram reveals a periodicity of 7.420$\pm$0.103 for the $g$-band and 7.394$\pm$0.826 day for the TESS light curve. In 2021, the peak periodicity of TESS light curves is $7.321\pm0.836$ days, similar to that in 2019. The $g$-band peak periodicity remains around $7.332\pm0.172$ days, but the Lomb-Scargle power for this peak decreases, and a new peak emerges at around 10 days, which is not seen from the periodograms of earlier $g$-band light curves. This change might be the result of elevated accretion activity in 2021.

\begin{figure}
    \centering
    \includegraphics[width=0.48\textwidth]{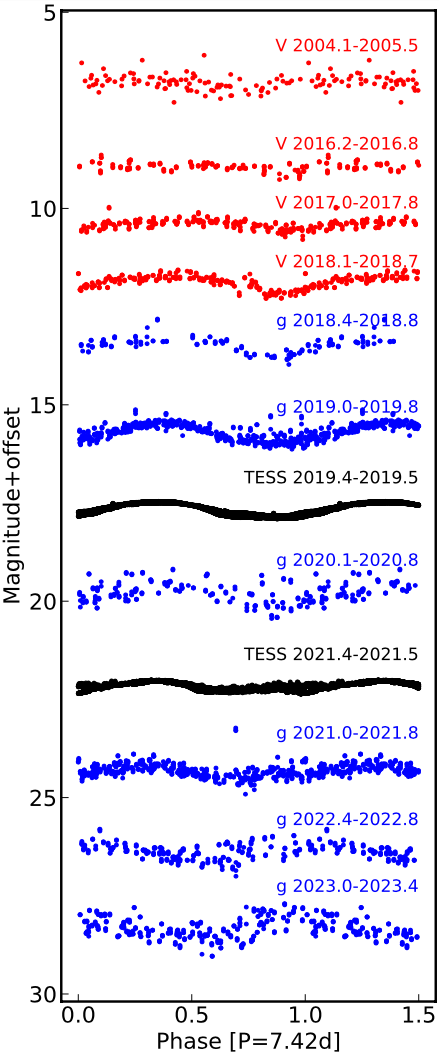}
    \caption{The phase-folded light curves of ASAS-SN $V-$ (red), $g-$band (blue),  and TESS (black) light curves, all folded to $P=7.42$ day and JD=2457200.  }
    \label{fig:asassn_phasefolded_lightcurve}
\end{figure}

Figure \ref{fig:asassn_phasefolded_lightcurve} shows the time evolution of spot modulation light curves. The ASAS-SN and TESS light curves between 2016 and 2023 are folded to $P=7.42$ day and JD=2457000. We also include a segment of $V$-band light curves taken between 2004 and 2005. The spot modulation profiles remain steady and with a small scatter between 2017 and 2019. After the 2020 burst, the $g$-band photometry brightens by 0.2 and the brightness minima has a larger scatter.  This larger scatter is consistent with the stochastic bursts seen at the minimum of modulation during the TESS observations in 2021 (Figure \ref{fig:tess}). After the 2022 burst, the modulation phase changed, and the original phases of brightness minima turned into maxima. We will further discuss this in Section \ref{sec:phase-lag-discussion}.

\subsection{The 2022 Burst}

\begin{figure}
    \centering
    \includegraphics[width= 0.45\textwidth]{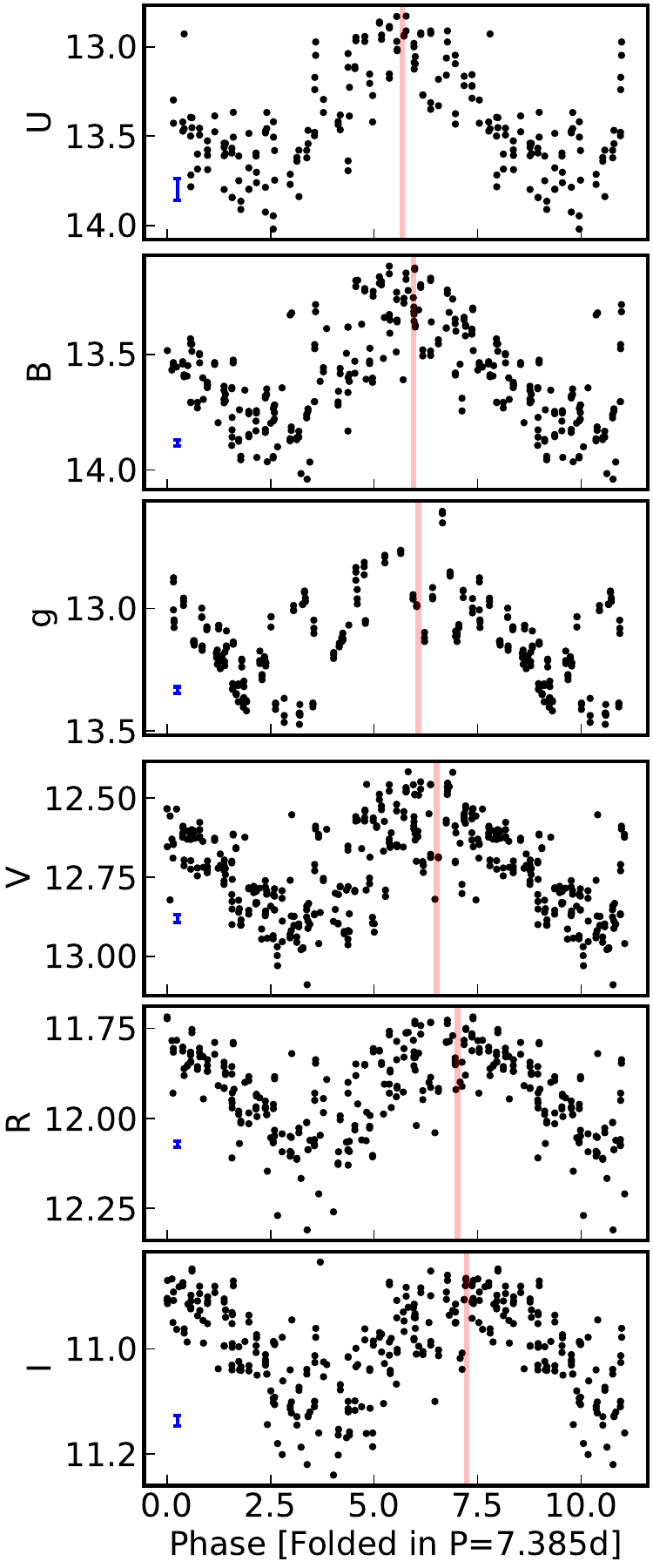}
    \caption{Phase-folded post-burst light curves in UBVRI and $g$ band from AAVSO. The multi-color light curves are phase-folded with P$=7.385$ day and to the U-band best-fit reference date JD$=2459716.96$. The red vertical lines indicate the location of brightness maxima for different wavelengths to show the phase lags. The blue errorbar at the lower left of each panel shows the median photometric errors in each band.}
    \label{fig:phasefolded_lightcurve}
\end{figure}

\begin{figure*}
    \centering
    \includegraphics[width=0.8\textwidth]{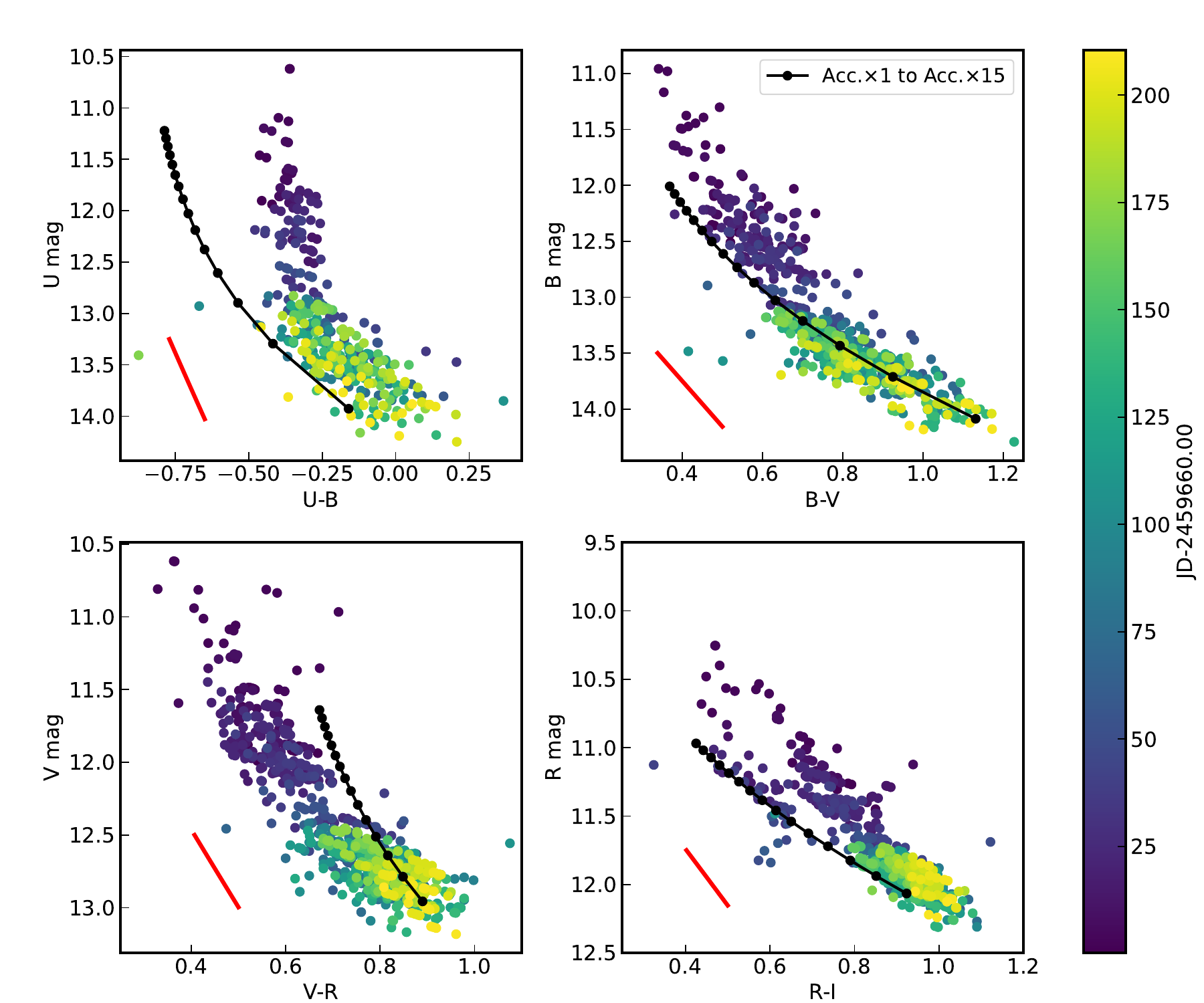}
    \caption{2022 burst on the color-magnitude diagram (CMD). The color is calculated using photometry taken at epochs separated by no more than one day. The data points are color-coded by the time since JD=2459600.0, the time when the burst reached its brightness maximum. The fainting of burst proceeded until JD=2459725.0, after which the variability is dominated by the spot modulations. The dotted black line is the color trend due to accretion enhancement, on the scale of one to fifteen times the accretion level inferred from the X-Shooter spectrum. The red line is interstellar reddening law of $A_V=0.5$ mag and $R_V=3.1$.}
    \label{fig:2022_burst_cmd}
\end{figure*}

EX Lup was reported to undergo a brightness rise from ASAS-SN $g$-band light curves early in 2022 \citep{zhou2022, kospal22}, triggering intense follow-up by astronomers who reported results to AAVSO.  The collection of multi-color light curves of the 2022 burst is displayed in the left panel of Figure \ref{fig:2022_burst_photometry}. 
On 2022-03-25 (MJD=59664.0), EX Lup reached its maximum, from where it faded until June 2022, when the $g$ photometry returned to pre-burst brightness. A tentative dip around 2022-08-08 (MJD=59800.0) and a slow fade after 2022-09-17 (MJD 59840.0) are also seen in light curves.

The following two subsubsections analyze the variability through periodicity seen through the burst and the changes in color versus brightness.  The periodicity during the burst is an accretion hot spot \citep[see also][]{Sicilia-Aguilar2015}, further modeled and discussed in Section \ref{sec:phase-lag-discussion}.  
The color changes versus brightness establish correlations with accretion rate, which we then apply to historical light curves.

\begin{figure}
\centering
    \includegraphics[width=0.45\textwidth]{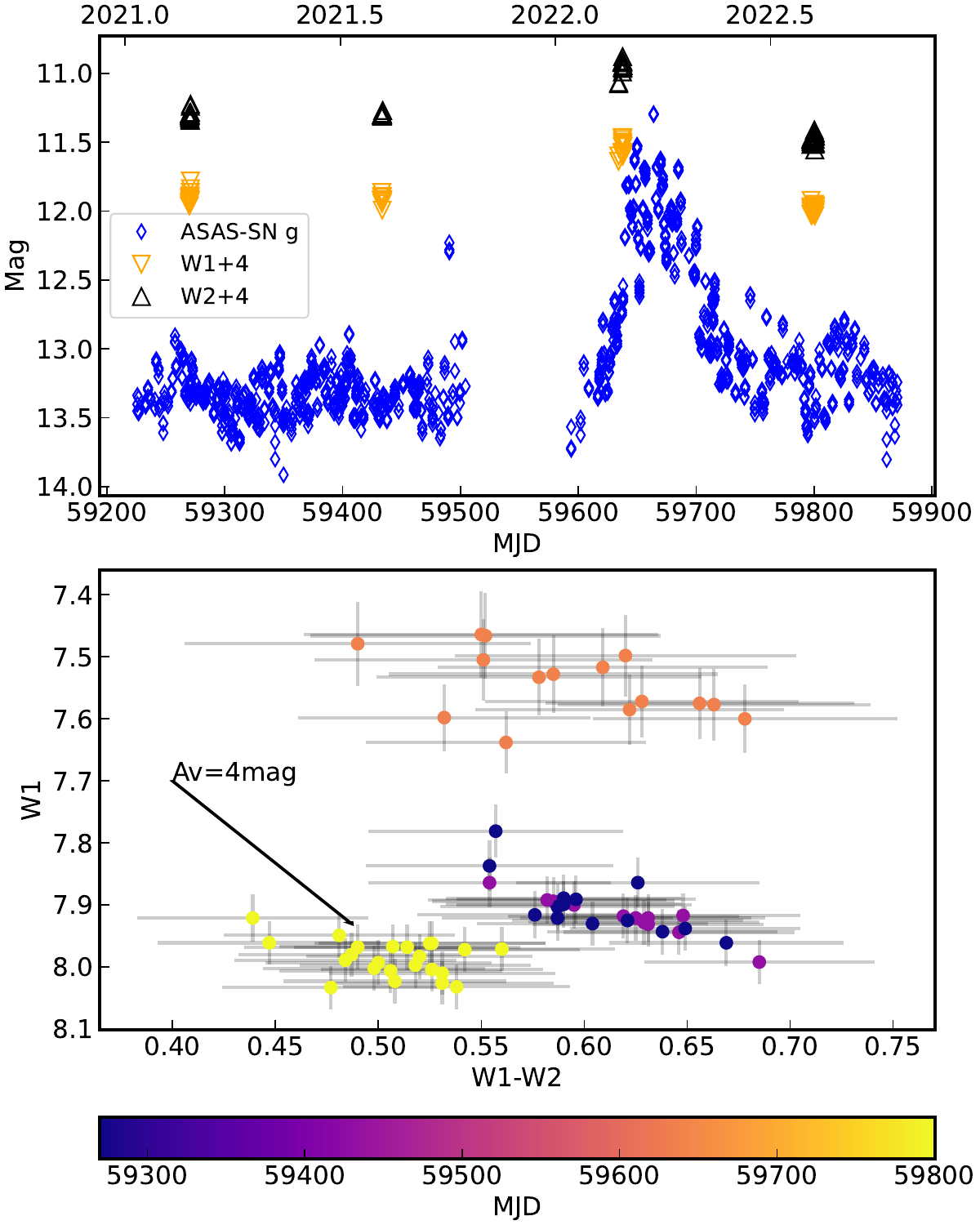}
\caption{Top: The ASAS-SN $g$ and NEOWISE W1/W2 photometry before and after the 2022 burst. Bottom: the color-magnitude diagram of W1 and W1-W2. The color trend predicted by the extinction of $A_V=4$ mag is indicated with a black arrow. }
    \label{fig:wise_cmd} 
\end{figure}

\subsubsection{Spot Modulation of the lightcurves}
\label{sec:spot}
 
Spot modulation is detected during and after the 2022 burst, as first reported during the rise by \citet{kospal22}. The post-burst light curves after 2022-05-25 (MJD 59725.0) show a strong periodicity peak in Lomb-Scargle Periodograms (right panel of Figure \ref{fig:2022_burst_photometry}). The in-burst light curves before 2022-05-25 (MJD 59725.0) also display spot modulations.

We fit the in-burst and post-burst light curves of each band with a sinusoidal function to determine the spot period ($P_{\rm sin}$), amplitude, and reference date ($T_0$).  Before this fit, we removed the fading trend with a linear fit (in magnitudes).
For the post-burst light curve, the fades in the light curves between MJD of 59780.0 and 59810.0 and after MJD=59840.0 are excluded.

The least-square fits of a sinusoid to the in-burst light curve yield a mean period of 7.55 days, which then decreases back to a mean of 7.385 days in post-burst light curves, both from multi-band fits (summarized in Table \ref{tab:spot_sin_fit}). We also include the peak periods  $P_{\rm LSP}$ from the Lomb-Scargle periodogram (LSP) in Table \ref{tab:spot_sin_fit}. The mean periods derived from in-burst light curves are 0.1 day longer than the periods measured during quiescence but within uncertainties; the longer measured periods are probably due to the larger photometric scatter and shorter observation span of in-burst photometry. 

The spot amplitudes derived from in-burst light curves range from 0.11 to 0.27 mag, while the amplitudes from post-burst light curves range from 0.1 to 0.33 mag. The in-burst spot amplitudes are generally smaller than those from post-burst. The amplitudes derived from shorter wavelengths are larger, as expected from the presence of cool or hot spots on the stellar surface \citep{herbst94}.  However, 
we measure differences in the reference date of the sinusoidal model for multi-color light curves, which suggests color-dependent phase lag in spot modulation between different wavelengths. To remove the bias of different periods, we re-fit the multi-color light curves but fix the periods at 7.55 days for in-burst light curves and 7.385 days for post-burst light curves, with reference dates listed in Table \ref{tab:spot_sin_fit}. 
Both in-burst and post-burst light curves show a longer lag towards longer wavelengths.
For example, the reference date difference between the U band and the I band in post-burst light curves is $\sim$ 1.5 days with a significance of 12.6 $\sigma$, while in the in-burst light curve, the $U-I$ phase lag is 1.25 days and has a lower significance of 5.3 $\sigma$.

To better visualize this phase lag, we present the phase-folded post-burst light curves in Figure \ref{fig:phasefolded_lightcurve}.
The phase lag between light curve peaks from different observing wavelengths is also reported by \cite{Kospal2014}, who used photometry from the optical V-band to near-infrared \emph{Spitzer} band, though that was only a tentative signature owing to the sparse photometry coverage. We will further discuss this phase lag feature in Section \ref{sec:phase-lag-discussion}.

\begin{deluxetable}{cCCCC}
\tablecaption{Sinusoidal Fits to Spot Modulations in 2022 Burst}
\tablewidth{0.5\textwidth} 
\label{tab:spot_sin_fit}
\tablehead{\colhead{Band} & \colhead{P$_{\rm LSP}$} & \colhead{P$_{\rm sin}$} & \colhead{Amplitude} & \colhead{$T_0$} \\ 
\colhead{} & \colhead{(day)} & \colhead{(day)} & \colhead{(mag)} & \colhead{(JD-2459000.0)} } 
\startdata
&&\mathrm{In-Burst}\tablenotemark{a}&\\
\hline
U	&	$7.53 \pm 0.58$	&	$7.54\pm0.14$	&	$0.27\pm0.05$	&	$717.08\pm0.19$	\\ 
B	&	$7.52 \pm 0.53$	&	$7.52\pm0.11$	&	$0.25\pm0.04$	&	$717.30\pm0.16$	\\ 
g	&	$7.57 \pm 0.51$	&	$7.51\pm0.11$	&	$0.27\pm0.04$	&	$717.29\pm0.18$	\\ 
V	&	$7.56 \pm 0.48$	&	$7.56\pm0.08$	&	$0.26\pm0.03$	&	$717.48\pm0.11$	\\ 
R	&	$7.62 \pm 0.51$	&	$7.52\pm0.08$	&	$0.18\pm0.02$	&	$717.71\pm0.13$	\\ 
I	&	$7.58 \pm 0.54$	&	$7.61\pm0.07$	&	$0.12\pm0.01$	&	$718.34\pm0.13$	\\ 
\hline
&&\rm{Post-Burst}\tablenotemark{b}&\\
\hline
U	&	$7.39 \pm 0.15$ &	$7.39\pm0.02$	& 	$0.33\pm0.03$	&	$717.08\pm0.10$	\\ 
B	&	$7.40 \pm 0.16$ &	$7.40\pm0.02$	& 	$0.25\pm0.02$	&	$717.34\pm0.08$	\\ 
g	&	$7.42 \pm 0.17$ &	$7.39\pm0.03$	& 	$0.18\pm0.03$	&	$717.45\pm0.16$	\\ 
V	&	$7.39 \pm 0.44$ &	$7.39\pm0.02$	& 	$0.14\pm0.01$	&	$717.89\pm0.08$	\\ 
R	&	$7.37 \pm 0.17$ &	$7.36\pm0.02$	& 	$0.12\pm0.01$	&	$718.40\pm0.09$	\\ 
I	&	$7.38 \pm 0.52$ &	$7.38\pm0.02$	& 	$0.10\pm0.01$	&	$718.63\pm0.07$	\\ 
\enddata
\tablecomments{{$P_{\rm sin}$ is the period determined from sinusoidal fits, and $P_{\rm LSP}$ is from Lomb-Scargle periodogram analysis}. The reference date $T_0$ is obtained from sinusoidal modeling by fixing P = 7.385 days for post-burst and P = 7.550 days for in-burst light curves.}
\tablenotetext{a}{Between MJD-59670.0 and MJD-59725.0}
\tablenotetext{b}{Between MJD-59725.0 and MJD-59875.0, but the segments between MJD-59780.0 and MJD-59810.0 is excluded.}
\end{deluxetable}

\subsubsection{Color-Magnitude Diagrams}
\label{sec: cmd}

The multi-color photometry during the fade of the 2022 burst is shown as color-magnitude diagrams in Figure \ref{fig:2022_burst_cmd}.
In general, EX Lup gets fainter because the accretion continuum is weaker while the spot-averaged photosphere does not change.  Since the accretion continuum is hotter and has a bluer color than the photosphere, EX Lup appears brighter at bluer colors.  
 However, during accretion bursts, the photosphere is $10-40$ times fainter than the accretion continuum in $U$ and $B$ bands and thus contributes little to the $U-B$ color.  The behavior during such bursts depends primarily on the shape of the accretion continuum, and thus change in $U-B$ is small as accretion decreases.

With the optical spectroscopy described in Section \ref{sec: optical_spec_quies}, we compare the observed color-magnitude trends with the ones synthesized from the accretion model fitted from X-Shooter spectra (see Section \ref{sec: optical_spec_quies}), marked by the black dotted line in Figure \ref{fig:2022_burst_cmd}. The predicted color trend is obtained by increasing the accretion fluxes from one to fifteen times that of the base, so that the predicted magnitude ranges are comparable to the observations. The predicted trends due to increasing accretion agree well with observed $B-V$, $V-R$, and $R-I$ color.  However, the observed $U-B$ colors are redder than predicted during the burst, constant in color at $U<13$, indicating that the in-burst color of accretion at short wavelengths is redder than that measured from the quiescent emission.  This $U-B$ behavior indicates that the Balmer jump is small, as expected when the accretion flow is optically thick (see for example the models of \citealt{calvet98}, where the optically thin pre-shock gas has a larger Balmer jump than the optically thick post-shock gas).

We also analyze the mid-infrared color behavior from WISE/NEOWISE, including epochs that cover before, during, and after the 2022 burst. The ASAS-SN $g$, the NEOWISE W1/W2 photometry, and the CMD are shown in Figure \ref{fig:wise_cmd}. In the first two epochs in 2021, EX Lup is quiescent and its location on the color-magnitude diagram is localized. In the epoch 20 days before the 2022 burst peak, W1 is 0.4 mag brighter with a similar W1-W2 color.  Fifty days after the end of the burst, EX Lup is fainter in both W1 and W2 bands and slightly bluer than the pre-burst color, indicating some minor change in either the inner disk or in contributions from the accretion flow. The trace of color variability does not follow the prediction of extinction events.

\section{Accretion Rates from Light Curves\label{sec:excess_emission_acc_rate}}

In this section, we develop methods to convert photometry into accretion rates.  We then calculate accretion rates from the past 10 years of high-quality monitoring, including the 2022 burst, and then assess accretion rates during the historical major outbursts and characteristic bursts.  Finally, we compare our accretion rates, obtained from excess accretion continuum emission, to accretion rates using other diagnostics and past accretion rates from the literature.

\subsection{Conversion Between Excess Emission and Accretion Luminosity\label{sec: vmag2Lacc}}

For EX Lup, as with most other accreting young stars, the accretion emission dominates over the photosphere at short wavelengths, especially at $\lambda < 4200$~\AA\ (see Figure~\ref{fig:exlup_spec} and also spectroscopic fits in, e.g., \citealt{pittman22}).
Among the most common photometric bands, the $U$-band best diagnoses the relative change in accretion level, although some correction for photospheric contribution is still needed.  We first calculate the accretion rate from $U$-band photometry and then use the $U$-band measurements to establish conversions for $B$, $g$, and $V$ bands.

The excess $U$-band flux due to accretion is converted to accretion luminosity using the relationship developed by \citet{gullbring98},
\begin{equation}
\label{eq:gullbring_acc_u}
\log(L_{\rm acc}/L_\odot) = 1.09^{+0.04}_{-0.18}~\log(L_{\rm U}/L_\odot) +0.98^{+0.02}_{-0.07}.
\end{equation}
The U-band luminosity of EX Lup, $L_{\rm U}$, is calculated as
\begin{equation}
    L_{\rm U} = 4\pi d^2 F_{\rm U, zero}\Delta \lambda_{U}10^{-0.4U_{\rm ex}},
\end{equation}
where $d=154.72$ pc is the distance of EX Lup, $F_{\rm U, zero}$ is the $U$ band zero magnitude flux, $\Delta \lambda_{U}=680$ \AA\  is the approximate width of the $U$ band filter, and $U_{\rm ex}$ is the $U$ band magnitude, corrected for extinction and after subtracting the underlying photospheric flux.

Leveraging the 2022 burst and post-burst multi-color photometry monitoring, we investigate how the accretion-induced excess emissions in the $U$ band are related to those in other wavelengths. In principle, spots, accretion variability, and dust extinction can all contribute to observed flux variations, so genuine excess flux from accretion should be isolated from other causes of variability. We subtract the in-burst and post-burst light curves with the best-fit spot models in Table \ref{tab:spot_sin_fit}. The two segments during the 2022 burst with signs of a slow fade and are also excluded from this calculation. 

We assume the observed fluxes consist only of contributions from accretion and the photosphere, so any observed flux above the photospheric level is caused by accretion alone. For the adopted data, the excess emission is calculated in magnitudes for different bands. For example, the excess emission in $U$ band $U_{\rm ex}$ is
\begin{equation}
    U_{\rm ex} = -2.5\log_{10}(10^{-0.4(U_{\rm obs}-A_U)}-10^{-0.4U_{\rm pho}}),
\end{equation}
where $U_{\rm obs}$ is the observed $U$ band magnitudes, $A_U=1.56~A_V$ is the extinction in Johnson $U$ band, and $U_{\rm pho}$ is the photosphere fluxes in the U band, synthesized from X-Shooter spectrum analysis (see Sec \ref{sec: optical_spec_quies}). 

\begin{figure*}
    \centering
    \includegraphics[width=1.0\textwidth]{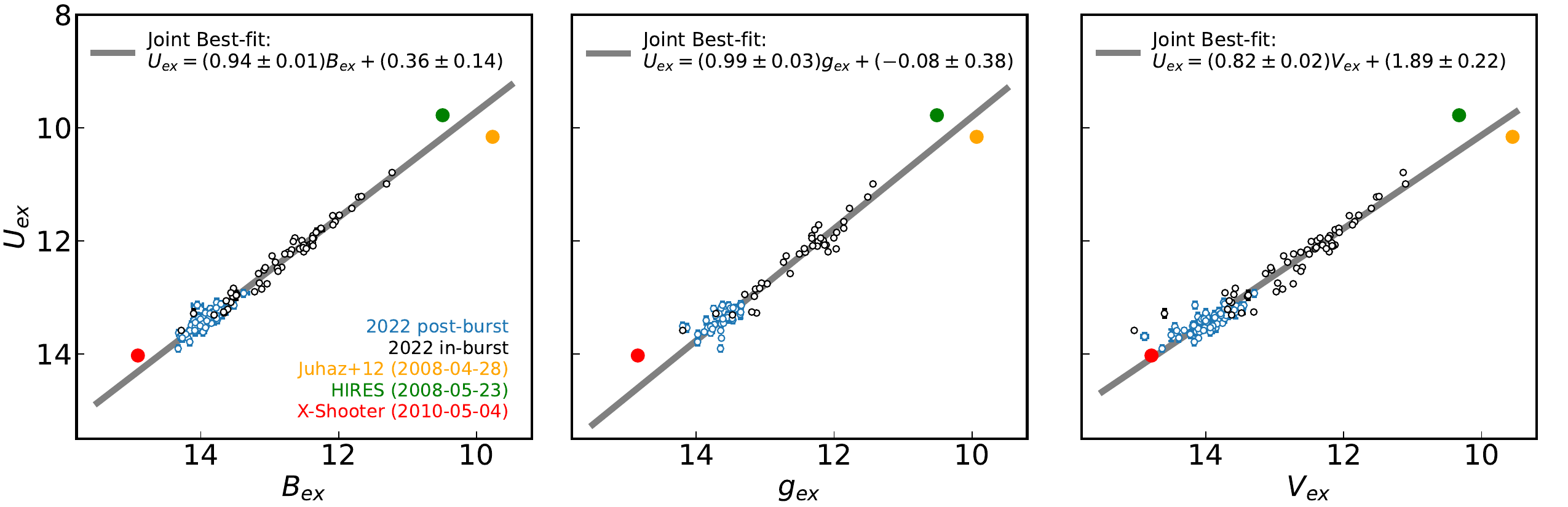}
    \caption{The relations of excess emission due to accretion in the U, B, g, and V band, derived from photometry and spectroscopy. Open black and blue dots are the excess emissions measured from 2022 in-burst and post-burst light curves, respectively. The gray line is the best-fit linear model to the 2022 burst excess emission measurements. For comparison, we also add excess emission measurements during the 2008 outburst from photometry \citep[yellow,][]{Juhasz2012} and spectrum \citep[green,][]{Aspin2010}, and during the quiescent state from X-Shooter spectrum (red, Figure \ref{fig:exlup_spec}).}
    \label{fig:excess_emission}
\end{figure*}

Following the procedures above, we calculate the excess emission in U, B, ASASN-SN $g$, and V band during the 2022 burst. The multi-color photometry is contemporaneous, with excess emission measurements separated by less than 0.5 days.
Figure \ref{fig:excess_emission} shows that the $B_{\rm ex}$, $g_{\rm ex}$, and $V_{\rm ex}$ are well correlated to the near-simultaneous $U_{\rm ex}$, with relationships\footnote{
The slopes $\Delta U_{\rm ex}/\Delta(BgV)_{\rm ex}$ fit to the post-burst data are smaller than those from in-burst data.  However, the measurements from the in-burst light curves span a much larger range  ($\Delta m_{\rm}\sim 2-3$ mag, compared to ($\Delta m_{\rm}\sim 1$ mag) during quiescence, so the combined fit is adopted here as more reliable.}
to the post-burst, in-burst, and combined light curves described by best-fit lines in Table~\ref{tab:excess_emission_fitting_result}.

A comparison between our spectroscopic accretion rate measurements and the accretion rates independently calculated from photometry establishes that the photometry-based accretion rates may be extrapolated to the large outbursts.
Figure \ref{fig:excess_emission} shows that excess emissions from our photometric method are similar to those obtained from 
 synthetic photometry from 
the HIRES spectrum in May 2008, from UBV photometry from WFI \citep{Juhasz2012}, and from quiescence with the X-shooter spectrum, with an average difference of 0.2 dex in the accretion rates (see also Section \ref{sec:accrate_compare}).  
In general, these three sets of excess emissions are beyond the calibrated range from the 2022 burst but are in agreement with the extrapolation of the calibrated relation. Therefore, we conclude that the excess emission relations we derived from the 2022 burst are applicable to a wide range, from the quiescent accretion state to the far more luminous outburst events.

\begin{deluxetable}{llCCCC}
\tablecaption{Linear Fits to the Excess Emission}
\label{tab:excess_emission_fitting_result}
\tablehead{\colhead{} & \colhead{} & \colhead{Slope} & \colhead{Intercept} & \colhead{rms [mag]} }\startdata
                        &   In-burst    & $0.93\pm0.02$	&	$0.42\pm0.25$	&	0.11      \\
$U_{\rm ex}-B_{\rm ex}$ &   Post-burst  & $0.80\pm0.06$	&	$2.26\pm0.79$	&	0.09 \\&   Joint-Fit   & $0.94\pm0.01$	&	$0.36\pm0.14$	&	0.10   \\
\hline                                    
                        &   In-burst    & $1.00\pm0.05$	&	$-0.24\pm0.65$  &	0.18   \\
$U_{\rm ex}-g_{\rm ex}$ &   Post-burst  & $0.57\pm0.13$	&	$5.56\pm1.76$	&	0.16 \\&   Joint-Fit   & $0.99\pm0.03$	&	$-0.09\pm0.38$  &	0.18 	    \\
\hline                                    
                        &   In-burst    & $0.81\pm0.03$	&	$2.03\pm0.39$	&	0.21    \\
$U_{\rm ex}-V_{\rm ex}$ &   Post-burst  & $0.50\pm0.04$	&	$6.37\pm0.61$	&	0.11 \\&   Joint-Fit   & $0.82\pm0.02$	&	$1.89\pm0.22$	&	0.18   \\
\enddata
\tablecomments{The lower limit of excess emission relation: $B_{\rm ex,low}$=14.9 mag, $g_{\rm ex,low}$=14.79 mag, $V_{\rm ex,low}$=14.83 mag.}
\end{deluxetable}

\subsection{Calculating accretion Rates from 2022 Burst \label{sec:accrate_var}}

Mass accretion rates $\dot{M}_{\rm acc}$ can be calculated as
\begin{equation}
\label{eq: macc_cal}
    \dot{M}_{\rm acc} = (1-\frac{R_*}{R_t})^{-1} \frac{L_{\rm acc}R_*}{GM_*}
\end{equation}
where $M_*=0.83~M_\odot$ and $R_*=1.50~R_\odot$ are the adopted mass and radius of EX Lup, the truncation radius $R_t$ is assumed to be $5$~$R_*$ \citep[adopted for consistency with other accretion rate measurements, see][]{hartmann16}\footnote{The truncation radius is typically assumed
$5$~$R_*$, based on the balance between the gas pressure and magnetic pressure \citep[e.g.][]{johnstone14}.  However, the truncation radius should depend on $\dot{M}^{2/7}$ \citep[e.g.][]{johnskrull07}, so epochs with larger accretion rates may have a smaller truncation radius.}, and the accretion luminosity $L_{\rm acc}$ is calculated from the U-band excess luminosity from Eq. \ref{eq:gullbring_acc_u} or from the excess fluxes in B, g, and V band through the relations in Table \ref{tab:excess_emission_fitting_result}.

\begin{figure}
    \centering
    \includegraphics[width=0.4\textwidth]{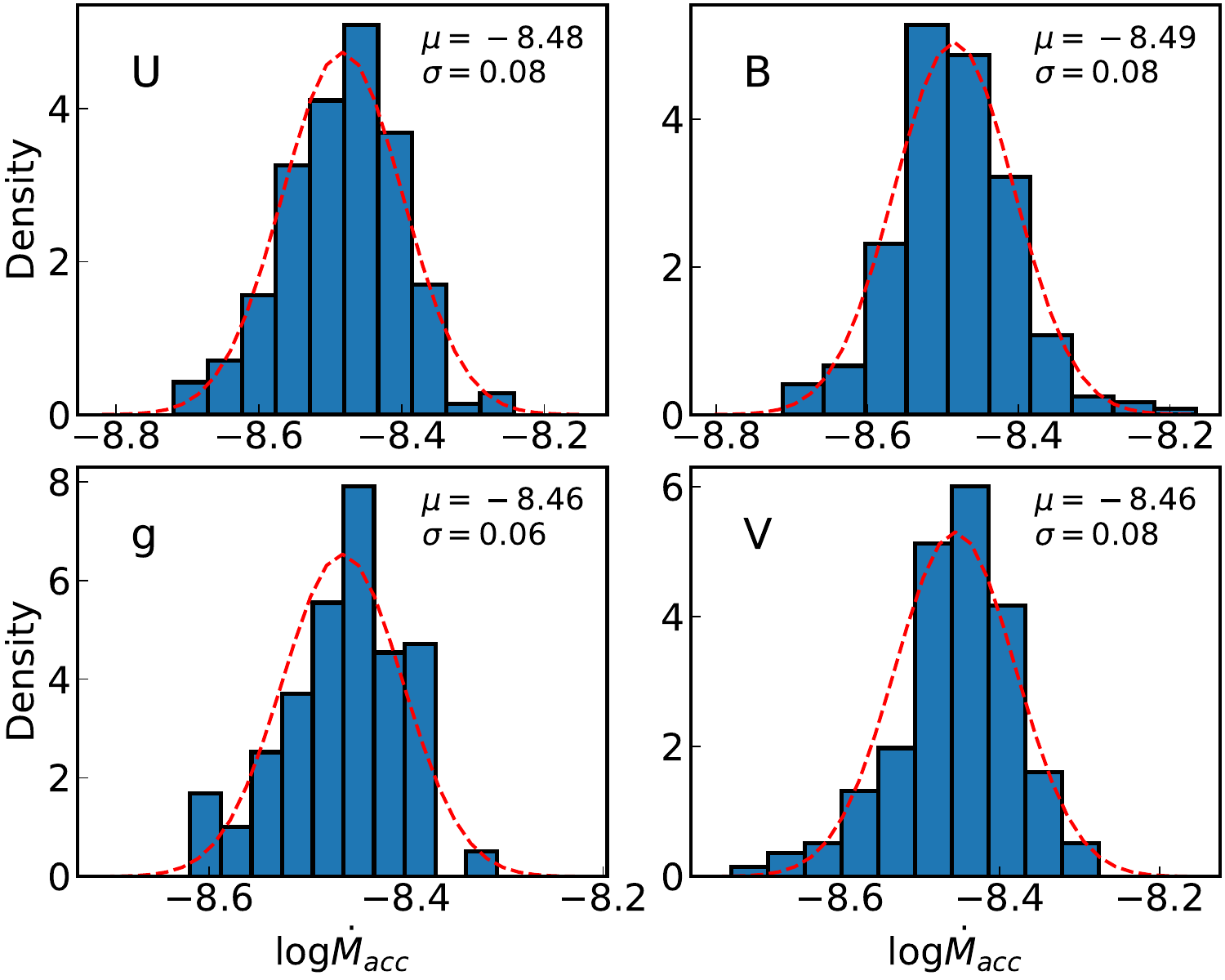}
    \caption{The histograms of mass accretion rates measured from the excess U, B, ASAS-SN g, and V band emission in the post-burst light curves of the 2022 burst. Gaussian fits the $\log \dot{M}_{\rm acc}$ histograms are plotted with red dashed lines.}
    \label{fig:2022_burst_maccrate_hist}
\end{figure}

Figure \ref{fig:2022_burst_maccrate_hist} presents the histograms of accretion rates measured from the excess emission of $U$, $B$, $g$, and $V$ band during the 2022 post-burst light curves. The two dips in brightness are excluded. The photometric variations induced by spot modulations are removed by the sinusoidal models (as listed in Table \ref{tab:spot_sin_fit}). Observations separated by less than two hours are binned into a single epoch. We also fit the Gaussian probability density functions to the histograms and show the best-fit mean value and 1$\sigma$ variance of the Gaussian function on the upper right of each panel. 

After the 2022 burst finished, the excess emission measurements from $g$, $B$, and $V$ all reveal a similar mean accretion rate of $3.33\times10^{-9}$ \mdotyr, two times 
 stronger than the accretion rate measurement of $1.77\times10^{-9}$\mdotyr\ in the X-Shooter spectrum in 2010. This increase in accretion rate quantifies what is seen visually, the steady brightening in the $g$-band of 0.6 mag from 2019--2023 (see Fig.~\ref{fig:assasn_wise}).
 The peak accretion rate of the 2022 burst is $1.74\times10^{-8}$ \mdotyr, as measured from U-band excess, which is $\sim5$ times higher than the post-burst accretion rates and $\sim 10$ times higher than the accretion rate prior to the burst.

The Gaussian sigma spread of post-burst accretion rates is around 0.08 dex, lower than the stochastic accretion variability measured for other accreting young stars (for example, see the analysis of a large sample by \citealt{venuti21} and in-depth analyses of individual stars by, e.g., \citealt{Zsidi2022} and Herczeg et al.~2023).
This distribution should be attributed to the day-to-month intrinsic accretion variability, due to the fact that the median cadence of post-burst observations is 1-2 days and the timespan of light curves is around 4 months, and we have removed the photometric modulations caused by spots. 
In the post-burst light curves, the rotational modulations induce larger photometric variations than the accretion variability.

If the rotational modulations were not corrected, the spread of the accretion rate derived from apparent photometry would increase to 0.14-0.22 dex.
Assuming the intrinsic accretion variability and modulation are independent, the intrinsic accretion variability only makes up 36-66\% of the apparent accretion rate spread. This agrees with the finding 
of \cite{Venuti2014} that rotational modulation of spots usually drives more significant photometric variations than intrinsic accretion variability on the timescales of the rotation period, at least for some objects.  

\begin{figure}
    \centering
    \includegraphics[width=0.4\textwidth]{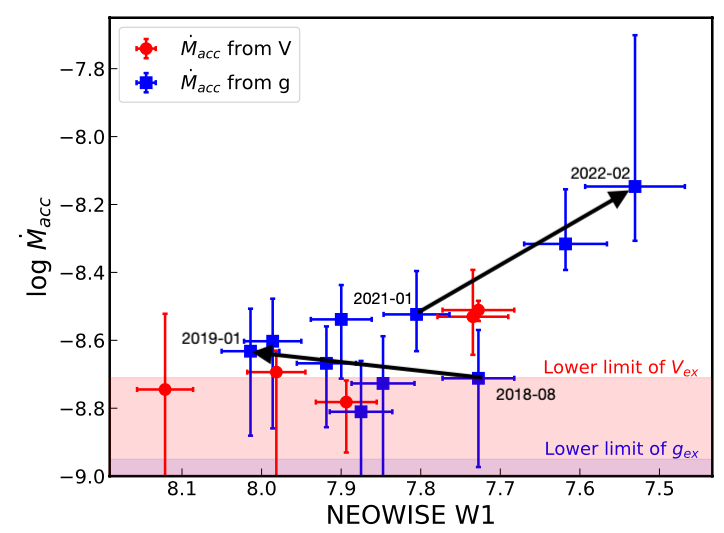}
    \caption{The relations between NEOWISE W1 photometry and mass accretion rates $\dot{M}_{\rm acc}$. The median mass accretion rates are derived from ASAS-SN $g$(blue) and $V$(red) band photometry taken within 4 days apart from the NEOWISE observation epochs, and the errorbars are the variability range of mass accretion rates in the 8-day time bins. The filled areas are the extrapolation ranges of our excess emission relations for $g$ and $V$ fluxes. }
    \label{fig:w1_maccrate}
\end{figure}

Figure \ref{fig:w1_maccrate} shows the relation between NEOWISE W1 photometry and the mass accretion rates of EX Lup.  The brightening of W1 tends to correlate with the increase in mass accretion rates. The brightest mid-IR points correspond to 2022-02, the rising stage of the 2022 burst, when the mid-IR emission is 0.35 mag brighter than quiescence while the accretion rate is 0.4 dex brighter than the pre-burst epoch, 2021-02.  However, in some epochs the $W1$ and $W2$ bands brightened without an accompanying change in accretion rate.  For example, albeit the $g$-band light curves show a 0.1 dex increase in mass accretion rates between epochs in 2018-08 and 2019-02, the W1 band faded by 0.3 mag from 2018 to 2019. Mid-IR emission is produced by the warm dust in the disk, although the exact interpretation of this emission and variability is challenging \citep[e.g.][]{mcginnis15,rebull15}.  The dominant emission should take the form of the irradiated disk surface (except in more extreme cases where viscous heating dominates, e.g., \citealt{rodriguez22}). The correlations between mass accretion rates and mid-infrared emission exhibited by EX Lup should help efforts \citep[e.g.][]{contreras20,contreras23} to calibrate extensive analyses of long-term mid-IR variability \citep[e.g.][]{park21,zakri22}.

\subsection{Accretion Rates During Major Outbursts and Characteristic Bursts}

\begin{figure*}
    \centering
    \begin{subfigure}
        \centering
        \includegraphics[width=0.9\textwidth]{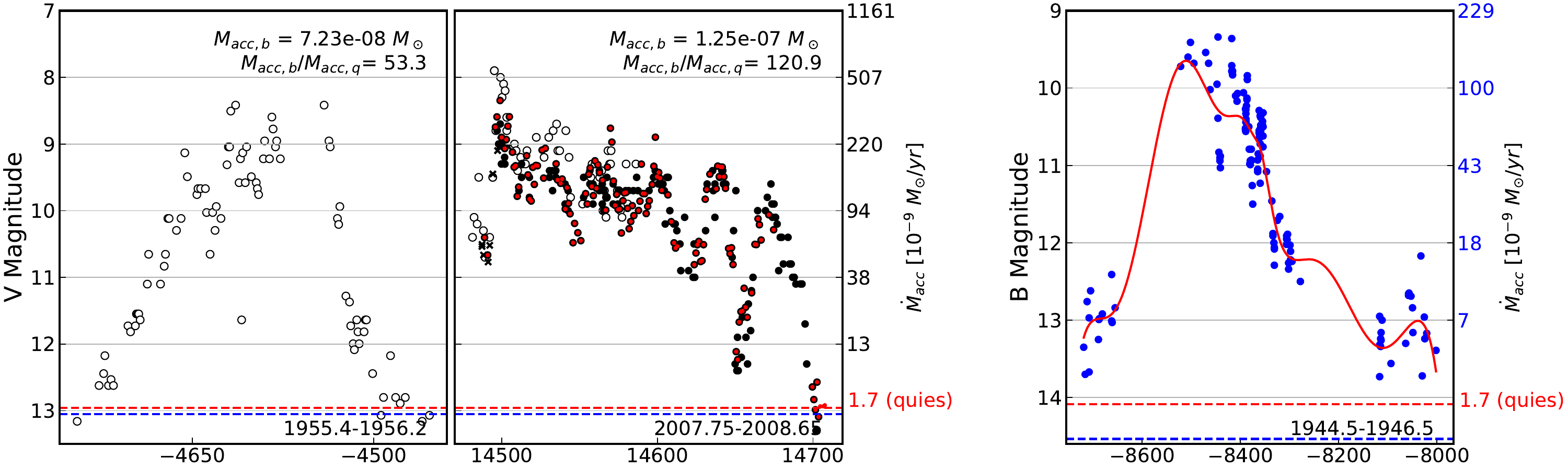}
    \end{subfigure}
    ~ 
    \begin{subfigure}
        \centering
        \includegraphics[width=0.9\textwidth]{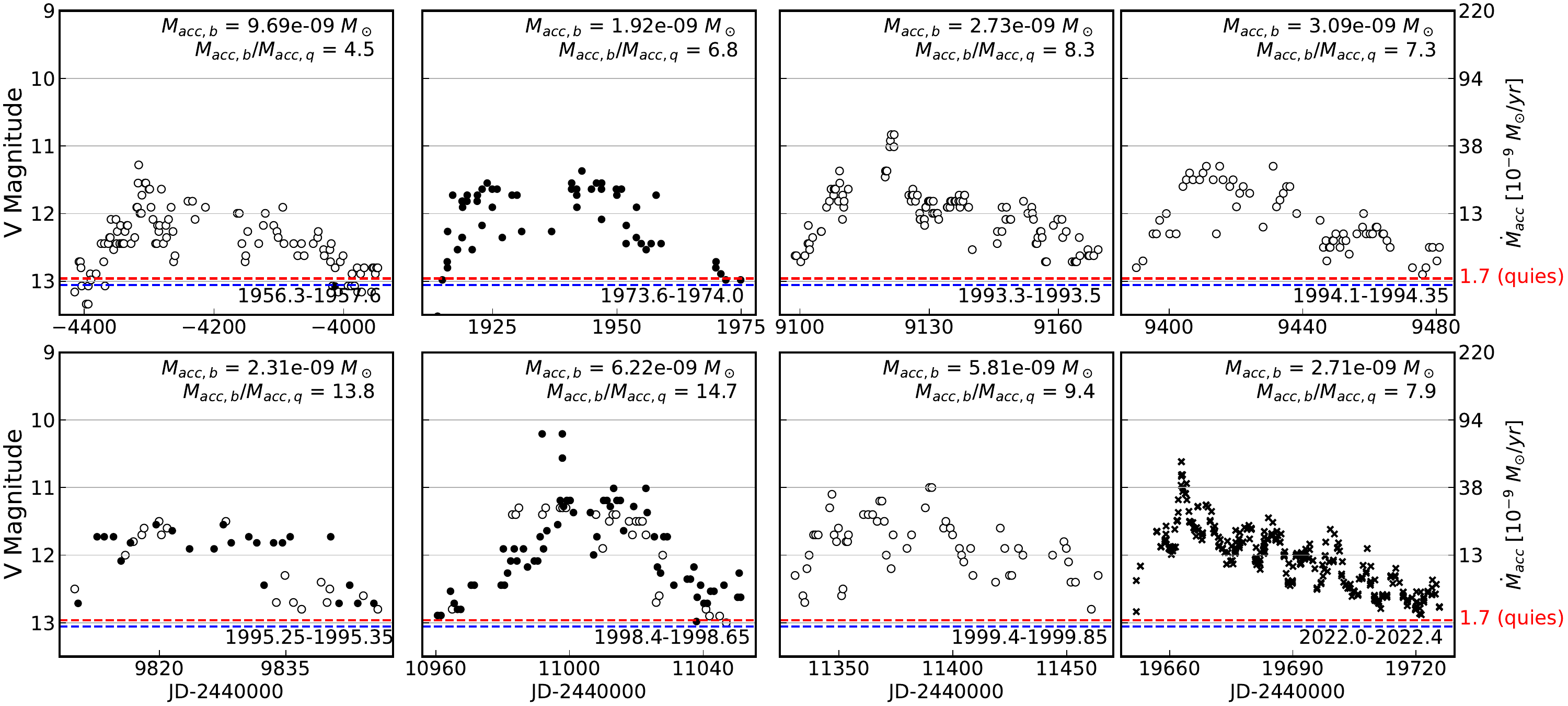}
    \end{subfigure}
    \caption{A selection of historical bursts in EX Lup. The three figures in the first row show EX Lup's historical outbursts in 1945, 1955, and 2008. The 1945 outburst is additionally fitted with a spline in red. The figures in the second and third rows show the characteristic bursts. The blue points are $B$-band magnitudes from DASCH light curves. The open circles are visual photometry from Albert Jones and are transformed to the Johnson $V$ band via Eq. \ref{eq:vis2v}. The black points are transformed AAVSO visual magnitude, and the black crosses are photometry made in the Johnson $V$ band from AAVSO. Red points are SMARTS $V$-band light curves. The equivalent mass accretion rates derived from excess emission relation are displayed on the right axes. The estimated accreted mass $M_{\rm acc,b}$ during each burst and its ratio to the quiescent accretion $M_{\rm acc,q}$ with the same duration is displayed on the upper right in each panel. The red dashed line stands for the V magnitude baseline for the mean quiescent rate ($1.7\times10^{-9}$\mdotyr). The blue dashed line is the $V$-band fluxes of EX Lup's photosphere. }
    \label{fig:good_burst_overview}
\end{figure*}

\begin{figure*}
    \centering
    \includegraphics[width=1.0\textwidth]{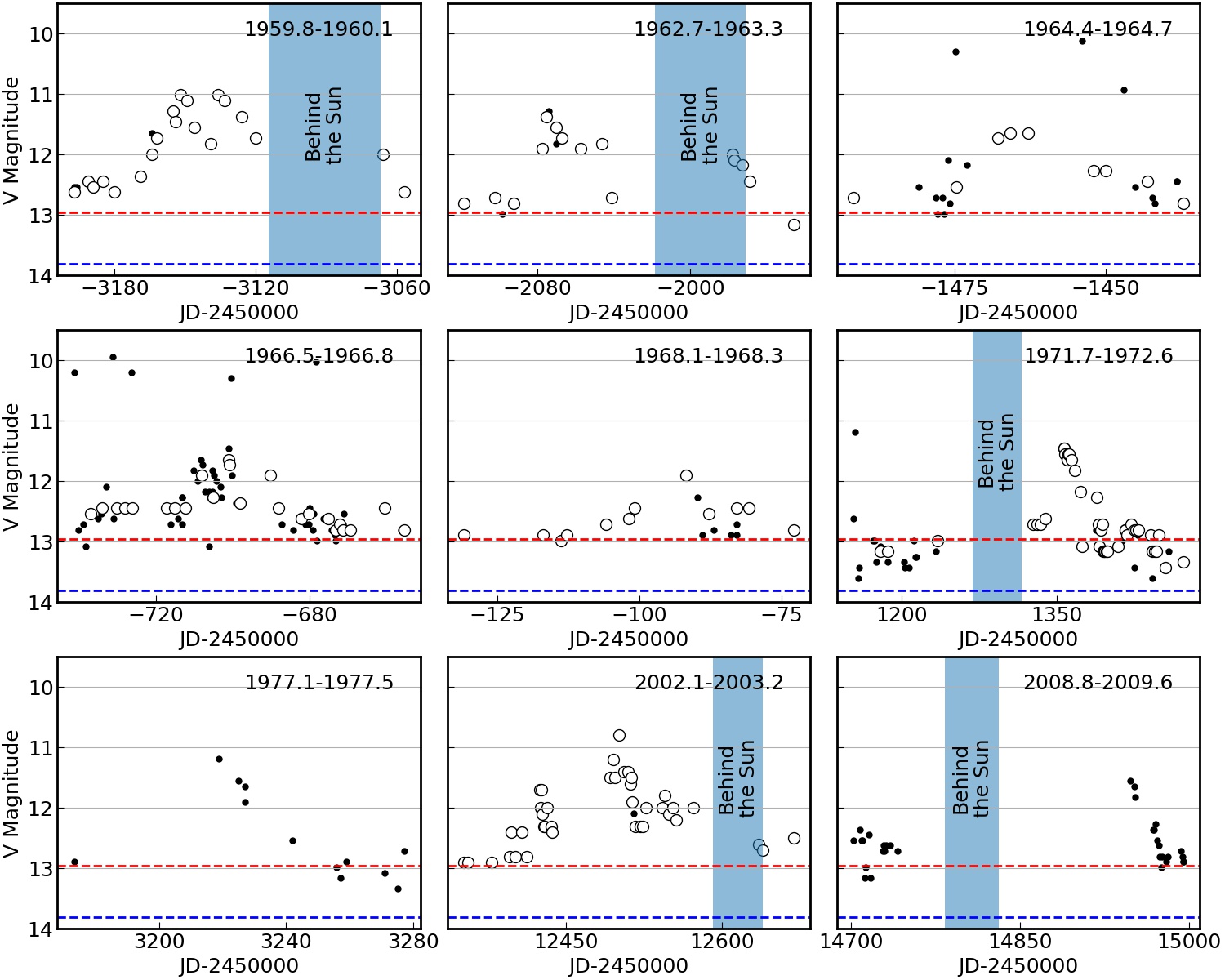}
    \caption{An overview of less characterized bursts seen from EX Lup. The photometry symbol scheme is the same as Figure \ref{fig:good_burst_overview}. These bursts are hard to have robust accreted masses estimates, either due to incomplete photometry coverage caused by seasonal observational gaps or inconsistent data reported by different observers. }
    \label{fig:bad_burst_overview}
\end{figure*}

A visual inspection is performed to identify the bursts from EX Lup historical light curves between 1945 to 2022. We first select a subset of ten bursts from AAVSO light curves that are well characterized from the photometric data (Figure \ref{fig:good_burst_overview}). These bursts have near-complete photometric coverage, with well-determined burst durations set by the time when EX Lup is continuously brighter than its quiescent state. We also include an outburst event at 1945, though with a poorly estimated burst duration, identified from DASCH light curves in Figure \ref{fig:good_burst_overview} to compare with the other two outburst events.
For completeness, we also include some less characterized bursts interrupted by long data gaps or inconsistent photometry data from different AAVSO observers, for which we cannot derive reliable $M_{\rm acc}$ from photometry. The light curves for this set of poorly characterized bursts are displayed in Figure \ref{fig:bad_burst_overview}.

The complete information of each burst, including the $V$ band magnitude and accretion rates at peak, burst duration, and total accreted mass, is tabulated in Table \ref{tab:burst_info}. We measure the mass accretion rates and accreted masses during the burst for these ten bursts by converting the observed visible magnitude to the Johnson V band using Eq. \ref{eq:vis2v}. Then accretion rates are calculated via the excess emission relations in Table \ref{tab:excess_emission_fitting_result}. The total accreted mass in the burst $M_{\rm acc,~b}$ is obtained by integrating all the photometry-derived mass accretion rates over the burst duration. We also calculate the ratio of $M_{\rm acc,~b}$ to quiescently accreted mass assuming the same duration $M_{\rm acc,~q}$ to provide a sense of the extra masses accreted during the burst. 

The major outbursts in 1955 and 2008 have much longer durations (hundreds of days) and reach $m_V\sim 8$ mag at peak. The peak mass accretion rate during these major eruptions is $\sim 3-5\times10^{-7}$ \mdotyr, hundreds of times larger than the average accretion rate during quiescence.  Around $0.1$ Earth masses can be accreted during these outbursts, which would otherwise take several decades to consume in the quiescent state.   The characteristic bursts have accretion rates of $2-5\times10^{-8}$ \mdotyr, ten times lower than the major outbursts.

The 1945 outburst does not have as complete photometric coverage as the other two outbursts, with photometry during only the fading stage spanning $\sim$200 days. Nevertheless, we put constraints on the mass accretion parameters by using a spline model fitted to longer segments of light curves to simulate the whole profile of the outburst, and by only using the in-fade photometry. These two methods estimate the upper and lower limit of accreted mass of the 1945 burst, respectively. The lower limit of peak mass accretion rate of the 1945 outburst is $~2.3\times10^{-7}$ \mdotyr, the magnitude of which is consistent with the 1955 and 2008 outbursts. The accreted mass is 6.2 to 8.9$\times10^{-8}~M_\odot$, similar to the 1955 outburst, but still lower than the 2008 outburst.

We also identify two long-term brightening events in DASCH light curves in 1937 and 1947 (see top panel of Figure \ref{fig:full_lightcurve}), each with a duration of 3-4 years, much longer than the typical duration of bursts (a few months). Taking the B magnitude calibrated by the APASS catalog and the excess emission relation in Table \ref{tab:excess_emission_fitting_result}, these two events brightened $\sim1$ mag in the B band, translating to a relative change in $\dot{M}_{\rm acc}$ of 0.47 dex and a peak accretion rate of $1.8\times10^{-8}$ \mdotyr. 
This accretion rate is 5 times larger than the mean accretion rates derived from the recent light curves and is close to the peak accretion rate in the 2022 burst.  

\begin{deluxetable*}{lcccccc}
\tablecaption{The Historical Burst Properties of EX Lup during 1945-2022}
\label{tab:burst_info}
\tablehead{\colhead{Burst Epoch} & \colhead{Duration} & \colhead{Peak Magnitude$^1$} & \colhead{Peak $\dot{M}_{\rm acc}$} & \colhead{$M_{\rm acc,b}$} & \colhead{$M_{\rm acc,b}/M_{\rm acc,q}$} & \colhead{Reference}  \\
\colhead{} & \colhead{(day)} & \colhead{(V band)} & \colhead{($\times10^{-9}\mdotyr$)} & \colhead{($\times10^{-9}M_\odot$)} & \colhead{} & \colhead{}  } 
\startdata
1945        & 250-500   &   9.5($B$)&   >232.94 &   62-89   &   43-54   &   -   \\  
1955 Apr.  & 350       &	8.3		&	358.84	&	72.32	&	53.34	&  a,f  \\
1956 Mar.   & 400       &	11.2	&	28.83	&	9.69	&	4.45	&  f    \\
1973 Aug.   & 80        &	11.3	&	26.34	&	1.92	&	6.77	&  e    \\
1993 Apr.   & 60        &	10.7	&	44.39	&	2.73	&	8.33	&  -    \\
1994 Jan.   & 80        &	11.2	&	28.27	&	3.09	&	7.32	&  b    \\
1995 Apr.   & 90        &	11.4	&	23.02	&	2.31	&	13.80	& a,b,d \\
1998 Jun.   & 110       &	11.0	&	38.02	&	6.22	&	14.74	& c     \\
1999 Jun.   & 80        &	10.9	&	37.92	&	5.81	&	9.39	& c     \\
2008 Jan.   & 200       &	7.8		&	551.37	&	125.14	&	120.87	&  c    \\
2022 Mar.  & 70        &	10.9	&	45.67	&	2.71	&	7.88	&  a    \\
\hline
1959 Sep.    & $>60^3$   & 11.1      &   -   & -        & -      & - \\
1962 Aug.    & $>80^3$   & 11.2      &   -   & -        & -      & - \\
1964 May     & 40$^3$    & 11.6$^4$  &   -   & -        & -      & - \\
1966 May     & 100$^4$   & 11.7      &   -   & -        & -      & - \\
1968 Jan.    & 50$^3$    & 11.9      &   -   & -        & -      & - \\
1972 Jan.    & $<100^3$  & 11.6      &   -   & -        & -      & - \\
1977 Feb.   & $<120^3$  & 10.65     &   -   & -        & -      & - \\
2002 Jul.   & $<210^3$  & 10.3      &   -   & -        & -      & c \\
2013        & $<500$    & $>11.0$   &   -   & -        & -      & - \\
2020        & -         & $>11.0(g)$   &   -   & -        &   -    & - \\
\enddata
\tablecomments{$M_{\rm acc,b}$ is the accreted mass during each burst, and $M_{\rm acc,q}$ is the accreted mass in the same duration of each burst and assuming quiescent accretion rate of $\dot{M}_{\rm acc,q}=1.7\times 10^{-9}$ \mdotyr ~(see Section \ref{sec: optical_spec_quies}). }
\tablerefs{a: \cite{Aspin2010}; b: \cite{Herbig2001}; c: \cite{Herbig2007}; d: \cite{Lehmann1995}; e: \cite{kospal22}; f: \cite{bateson91}}
\tablenotetext{1}{Converted from Visual magnitude via Eq. \ref{eq:vis2v}.}
\tablenotetext{2}{Measured from Johnson V band.}
\tablenotetext{3}{Sparse photometry coverage.}
\tablenotetext{4}{Inconsistent visual magnitude between different observers.}
\end{deluxetable*}

\subsection{Comparisons to other accretion rate measurements for EX Lup \label{sec:accrate_compare}}

The accretion luminosities presented in this work are obtained by directly measuring the excess accretion continuum emission.  Table~\ref{tab:acc_comparison} compares our accretion rates with selected accretion rates from the literature and from other spectroscopic diagnostics.  The accretion rates presented in this work are much more likely to be underestimated than overestimated due to the exclusion of low-density flows (see \citealt{Espaillat2021}, \citealt{pittman22} and references therein) and the exclusion of line emission (see discussions in \citealt{alcala14} and Herczeg et al.~2023).

The Br$\gamma$ emission is detected in emission, weakly during quiescence and then stronger as the accretion rate increases.  The accretion rates inferred from the Br$\gamma$ luminosity are well correlated with the accretion rates from the continuum, lending some confidence to the application of global correlations between Br$\gamma$ line and continuum luminosity \citep[e.g.][]{alcala17} to EX Lup bursts and outbursts.  However, the Pa$\beta$ emission measured by \cite{Sipos2009} leads to an accretion rate that is a factor of 10 lower than the continuum-based accretion rates.

 The quiescent accretion rates are systematically larger than the accretion rates of $\sim 10^{-10}$ \mdotyr~that are estimated from H$\alpha$ line widths \citep[e.g.][]{Sicilia-Aguilar2012,Sicilia-Aguilar2015}.  The difference is likely attributed to different methodologies.  Excess continuum emission should be considered more reliable, since multiple factors affect the H$\alpha$ line profile \citep[see also, e.g.][]{manara17}.

\begin{table*}[!t]
\caption{Comparison of Accretion Rates Between Spectra and Photometry \label{tab:acc_comparison}}

\centering
\begin{tabular}{lcccccccc}
\toprule
                   &            &                   &      & \multicolumn{4}{c}{$\dot{M}_{\rm acc}$} &  \\
Instrument & Date & Accretion   & $\dot{M}_{\rm acc}$                  & \multicolumn{4}{c}{From Excess Emission$^7$}                 & Ref. \\
           &      & Diagnostics & ($\times 10^{-9}~M_\odot$ yr$^{-1}$) & \multicolumn{4}{c}{($\times 10^{-9}~M_\odot$ yr$^{-1}$)} &      \\
                   &            &                   &      & $U$      & $B$     & $g$      & $V$     &  \\ \midrule
Keck/HIRES         & 2008-05-23 & Continuum         & 150  & 124.02   & 79.50   & 72.45    & 69.40   & This work$^1$ \\
Keck/NIRSPEC       & 2008-05-23 & Br $\gamma$       & 140  & 124.02   & 79.50   & 72.45    & 69.40   & This work$^{2,3}$ \\
NTT/SOFI           & 2008-04-20 & Br $\gamma$       & 200  & 84.57    & 60.85   & 136.35   & 132.58  &  a$^{4,6}$\\
MPG/FEROS          & 2008-05-08 & H$\alpha$ 10\% EW & 30   & -        & -       & -        & 120     &  b$^5$\\
GeminiSouth/IGRINS & 2022-03-14 & Br $\gamma$       & 8.9  & -        & -       & 9.4      & 10.5    & This work$^2$\\
VLT/X-Shooter      & 2010-05-04 & Continuum         & 1.77 & 1.74     & 1.22    & 0.97     & 1.76    & This work$^1$ \\
VLT/X-Shooter      & 2010-05-04 & Br $\gamma$       & 1.34 & 1.74     & 1.22    & 0.97     & 1.76    & This work$^{1,2}$ \\
MPG/FEROS          & 2007-07-29 & H$\alpha$ 10\% EW & 0.3  & -        & -       & -        & 4.1     & b$^5$\\
NTT/SOFI           & 2001-05-04 & Pa$\beta$         & 0.4  & -        & -       & -        & 5.13    & c$^6$\\ \bottomrule
\multicolumn{9}{l}{a: \citet{Juhasz2012}; b: \citet{Sicilia-Aguilar2012}; c: \citet{Sipos2009}.} \\
\multicolumn{9}{l}{1. Synthetic magnitudes from the spectrum.}\\
\multicolumn{9}{l}{2. $L_{\rm line}-L_{\rm acc}$ from \citet{alcala17}.}\\
\multicolumn{9}{l}{3. Equivalent width from \citet{Aspin2010}, scaled with K-band photometry from \citep{Juhasz2012}.}\\
\multicolumn{9}{l}{4. Photometry from WFI observation \citep{Juhasz2012}.}\\
\multicolumn{9}{l}{5. $\dot{M}_{\rm acc}-EW_{H\alpha}$ from \citet{Natta04}}\\
\multicolumn{9}{l}{6. $L_{\rm line}-L_{\rm acc}$ from \citet{Muzerolle98}}\\
\multicolumn{9}{l}{7. Assuming $A_V=0.1$ mag, $A_U=1.5A_V$, $A_B=1.3A_V$, $A_g=1.2A_V$.}
\end{tabular}
\end{table*}

\section{Discussion\label{sec:historical_bursts}}

In this Discussion, we first provide an overview of the major outbursts and minor bursts detected over the past 130 years (Section \ref{sec:exlup_burst}).  This accretion history tells us that over the past 70 years, about two times more mass has accreted onto EX Lup during outbursts than in quiescence, as described in Section \ref{sec:quies_burst_accretedmasses_ratio}.  The description of the bursts and quiescence is used in Section \ref{sec:physics} to offer constraints and challenges to the two hypothetical explanations for EX Lup-type bursts.  We then describe in Section \ref{sec:phase-lag-discussion} the spotted surface of EX Lup and consequences for photometric variability, which needs to be considered for the accretion rates during quiescence.

\subsection{A Historical Overview of Accretion Bursts onto EX Lup \label{sec:exlup_burst}}

\begin{figure*}
    \centering
    \includegraphics[width=1.0\textwidth]{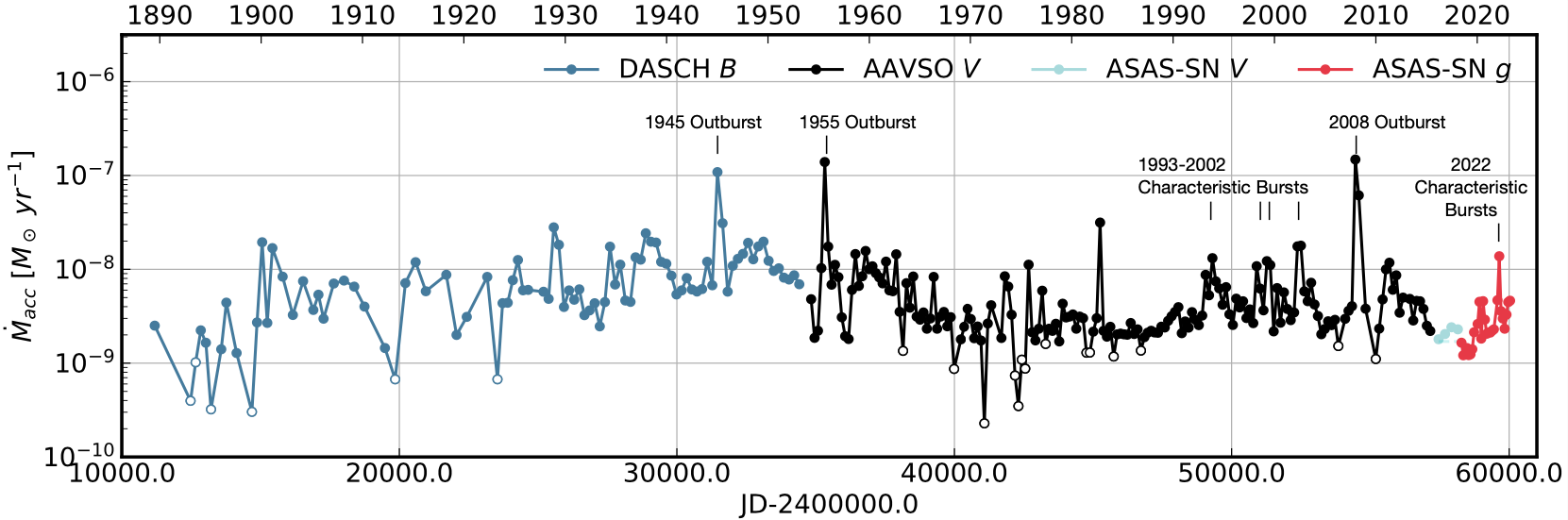}
    \caption{The mass accretion rates of EX Lup between 1890 and 2022. DASCH $B$-band, AAVSO $V$-band, and ASAS-SN $g$- and $V$- band light curves are used to derive the accretion rates. Filled and hollow points are within and beyond the calibrated ranges of excess emission relations, respectively. }
    \label{fig:accretionrate_overview}
\end{figure*}

Figure \ref{fig:accretionrate_overview} shows the historical light curve in accretion rate from 1893--2022. The accretion rates are derived from mean magnitudes in half-year bins for DASCH light curves, 100-day bins for AAVSO light curves, and 60-day bins for ASAS-SN light curves. The century-long accretion history shows a gradual decrease in accretion rate up to 1 dex from the 1940s to the 1980s. The large spikes in the accretion rate evolution are attributed to accretion bursts, which we further characterize in this subsection and evaluate their roles in EX Lup's historical accretion in Section \ref{sec:quies_burst_accretedmasses_ratio}.

The historical bursts from EX Lup are divided here into two categories: characteristic bursts and major outbursts. The characteristic bursts have a typical peak magnitude of $m_V\sim 11.5-10.5$ and last for $\lesssim$ 100 days. The mass accretion rate during these characteristic bursts is elevated above the quiescent level by 10--30 at their peak.
These characteristic bursts accrete masses $\sim 10^{-3}$ Earth masses, or around ten times the quiescently accreted mass during the same time period.
The recent 2022 burst also falls into the category of a characteristic burst.

\begin{figure}
    \centering
    \includegraphics[width=.48\textwidth]{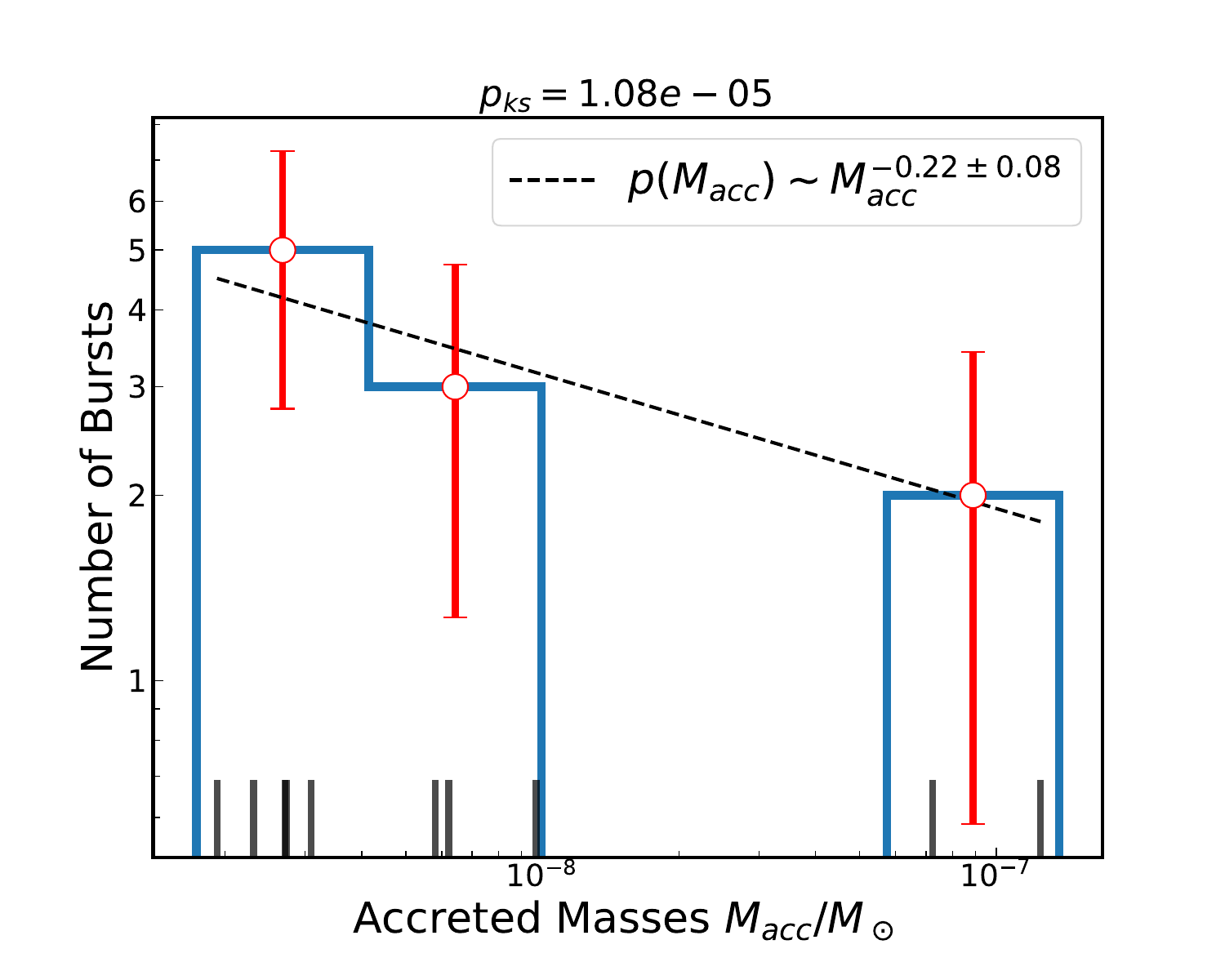}
    \caption{The histogram of accreted masses $M_{\rm acc}$ of ten bursts from Figure \ref{fig:good_burst_overview}. The black sticks at the bottom mark the value of $M_{\rm acc}$ for each burst. The error bars of the number of bursts in each $M_{\rm acc}$ bin are the Poisson error due to the small sample size. The dashed line is a power law relation fitted to the burst number in the three accreted masses bins. However, the KS test suggests the accreted mass of 10 bursts is not likely sampled from such power law distribution (with a p-value of $1.08\times10^{-5}$), due to the non-detections between $10^{-8.0}\sim10^{-7.25}~M_\odot$.}
    \label{fig:burst_histogram}
\end{figure}

Figure \ref{fig:burst_histogram} shows the histogram of $M_{\rm acc}$ of ten well-characterized bursts. Six equal-width bins are used to plot the histogram. Admittedly, the number of bursts is small and the choice of the bin will affect the statistics, but it generally reflects a trend that the characteristic bursts are more common in EX Lup. No bursts in the sample have $M_{\rm acc}$ between $10^{-8.0}\sim10^{-7.25}~M_\odot$.

To obtain the statistical significance for the lack of bursts in this intermediate $M_{\rm acc}$ bin, we fit a power law to the number of bursts versus the accreted masses. Then we draw samples from the fitted power law and perform the Kolmogorov-Smirnov test with our sample. The $p-$value is $1.08\times10^{-5}$, suggesting the accreted mass of bursts seen from EX Lup are inconsistent with a single power-law distribution. The two distinctive classes of bursts seen from EX Lup suggest that the driving causes of these bursts are not regulated by self-organized criticality \citep{bak87}.  In other words, the distribution of mass accreted in bursts form two distinct populations, the characteristic bursts, and outbursts, which might not share common driving mechanisms.

The adopted uncertainty of extinction and the excess emission relation introduces 15\% and 40\% uncertainties to the accretion rates and total accreted mass estimate, which is not sufficient to reconcile the gaps in accreted mass histogram. If the truncation radius shrinks during bursts, the accretion rate during bursts would be expected to be even higher than measured here.
The efforts of compiling a complete EX Lup bursts record would also fail due to the sporadic data gaps, so we cannot completely rule out bursts of intermediate size. However, the major reason for incompleteness is the $\sim 2-3$-month seasonal data gaps every year, comparable to the typical duration of characteristic bursts ($\sim 100$ days).

If the extinction is $A_V=1.1$ mag, as measured by \citet{alcala17} and adopted by Cruz S\'aenz de Miera et al.~(2023), all accretion rates would be roughly a factor of $\sim 6.5$ times higher than estimated here (see change in accretion luminosity and radius described in \S 3.1).

\subsection{The Ratio of Accreted Masses During Bursts and Quiescent State \label{sec:quies_burst_accretedmasses_ratio}}

We have now tabulated the history of accretion onto EX Lup over the past century, including a tabulation of bursts and accretion rates and an assessment of accretion during quiescent periods.  These accretion rates are calculated by converting historical photometry to accretion rate through correlations developed using the multi-color photometry in the 2022 burst.  To move forward, we ask the question: \emph{is more mass accreted by EX Lup during bursts or during quiescent?}

An exact comparison is difficult, owing to the non-uniform photometric data coverage in 130 years, especially because the sparse monitoring before 1955 affects our knowledge about the true frequency of bursts in EX Lup. As a consequence, we restrict the discussion to the 70-year photometry from AAVSO and ASAS-SN light curves, during which we have a complete census of major bursts and a near-complete census of characteristic bursts.  For the characteristic bursts, the photometry may not be as complete due to the seasonal data gaps (two months per year), therefore we estimate completeness of 84\% to correct for the characteristic burst frequency. 

As shown in Table \ref{tab:burst_info}. The major outbursts in 1955 and another in 2008 have a total (combined) mass accreted of $1.97\times10^{-7}~M_\odot$.
The mean accreted masses in characteristic bursts is $3.68\times10^{-9}~M_\odot$. Assuming the apparent frequency of characteristic bursts is two per decade, further compensating the 84\% completeness and plus two major outbursts, the accreted masses during bursts $M_{\rm acc,~b}$ between 1955 and 2022 is
\begin{equation}
\begin{split}
    M_{\rm acc,~b}= \frac{14}{0.84}\times3.68\times10^{-9} M_\odot + 1.97\times10^{-7} M_\odot     \\
    = 2.6\times10^{-7}~M_\odot.
    \end{split}
\end{equation}
Despite the higher frequency of characteristic bursts, the major outbursts are responsible for $\sim 80\%$ of masses accreted during all bursts. 

Taking the mass accretion rate during quiescence of $\dot{M}_{\rm acc,~q}=1.7\times10^{-9}$ \mdotyr, as measured in Section \ref{sec: optical_spec_quies}, a total of $1.1\times10^{-7}~M_\odot$ is accreted during quiescent between 1955 and 2022. 
Therefore, In this case, $M_{\rm acc,~q}/M_{\rm acc,~b} \sim 1:2$, the amount of mass accreted during the bursts is two times larger than the mass accreted in quiescence.  However, the accretion rate of EX Lup during the quiescence is variable. For example, in Section \ref{sec:accrate_var}, we found that EX Lup exhibited a $\sim0.2$ dex accretion rate increase between 2010 and 2022, as seen 
by the gradual brightening in ASAS-SN $g$ light curves (see Figure \ref{fig:assasn_wise} also). The DASCH light curves also exhibit two long-term brightening events around 1938 and 1949, both of which lasted 2-3 years.  The peak accretion rate of $1.8\times10^{-8}$ \mdotyr\ is similar to those measured from the AAVSO characteristic bursts, indicating that the quiescent accretion can also reach the $\dot{M}_{\rm acc}$ value that characteristics burst can attain at peak, but in a slower way.  These changes bring the quiescent and burst accretion totals closer to equality.

Since the outburst-consumed mass dominates the total accreted mass budget in all bursts, we caution that the ratio calculated above will be largely sensitive to the outburst frequency. The apparent outburst frequency in AAVSO 68-year data is one event per 34 years, which is similar to the outburst frequency between 1890-2022 (one event per 44 years), assuming the 1945 outburst is the only outburst present between 1890-1953. Therefore, we cannot draw conclusions on whether the outburst frequency is overestimated or not based on historical data.

We  conclude that roughly half of the mass of EX Lup accretes during major accretion bursts and half accretes during characteristic bursts and quiescence.
The quiescent accretion rate of 1.7-3.3$\times10^{-9}$ \mdotyr\ measured across the past ten years is similar to (slightly lower than) the accretion rate of $3\times10^{-9}$ \mdotyr\ expected for a 0.83 M$_\odot$ star, based on the relationship\footnote{scaled for Gaia DR3 distances, as in \citet{manara21}, and with masses rescaled to the evolutionary tracks with 50\% spots of \citet{somers20}.} for complete samples of disks in Lupus and Cham I from \citet{alcala17} and \citet{manara17}.  Incorporating the full accretion history, including periods of quiescence with an elevated steady accretion rate, increases the accretion rate of EX Lup by a factor of $\sim 4$, a factor of a few higher than expected for its mass (from a sample of single-epoch accretion rates). The ratio of gas accreted in bursts versus quiescence is robust to extinction uncertainty, since all accretion rates would be shifted by a similar amount.

\subsection{Mass loading in the innermost disk\label{sec:physics}}

The detailed accretion history of EX Lup should help to motivate improvements and constrain  models to understand these short outbursts. Accretion outbursts are generically explained by an excess of mass in the inner disk, leading to excess accretion until that mass has accreted or otherwise dispersed \citep[see, e.g.][]{hartmann96}.  The excess mass may be located only at the inner disk or may be spread across larger radii.

\begin{figure}
    \centering
    \includegraphics[width=.48\textwidth]{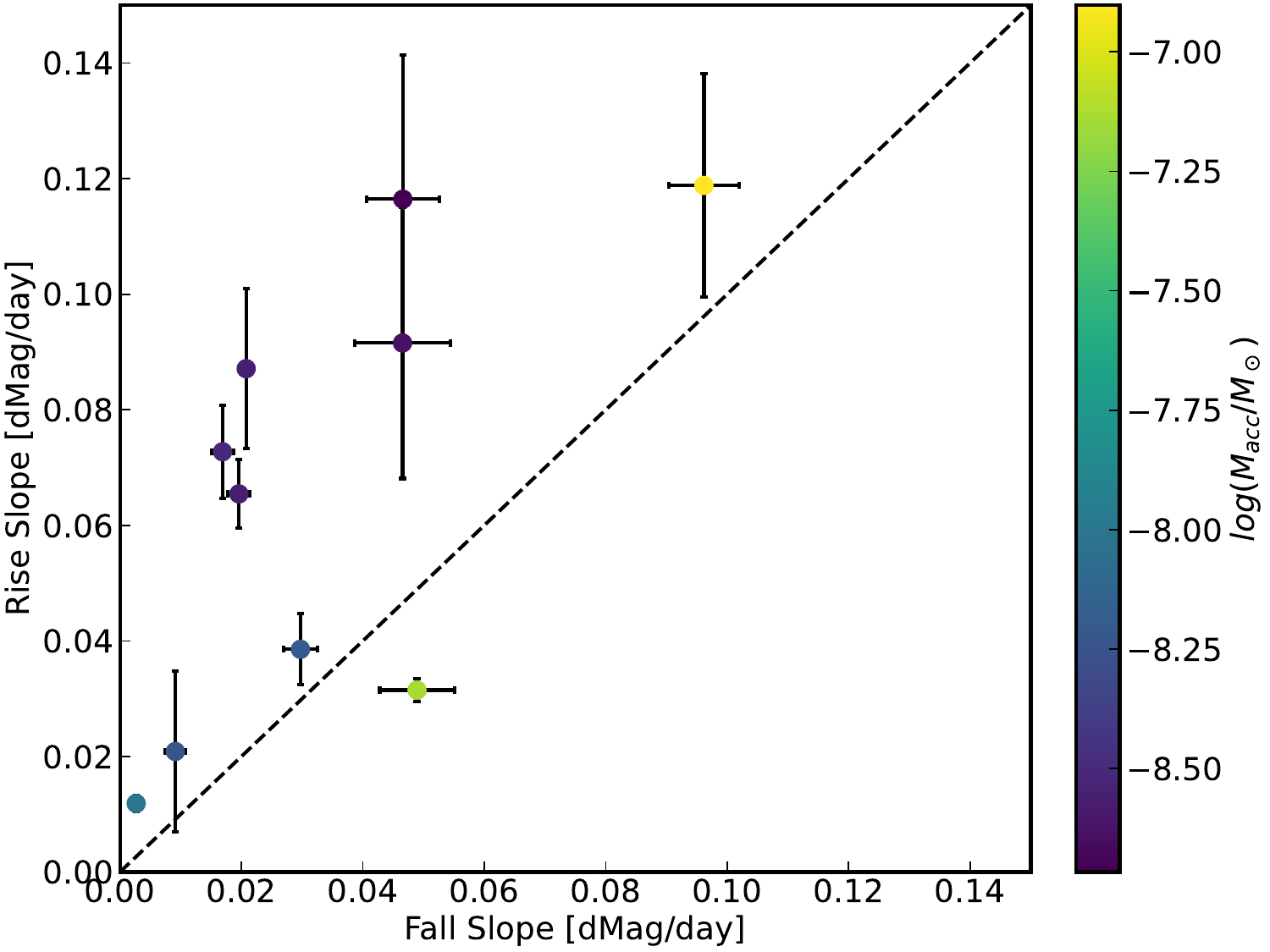}
    \caption{The rise and fall slopes in magnitudes per day, measured from ten bursts in Figure \ref{fig:good_burst_overview} (excluding the 1945 outburst), color-coded by the accreted mass during the burst. }
    \label{fig:burst_slope}
\end{figure}

For EX Lup, the outbursts are strictly an inner disk phenomenon, for two reasons.  First, as seen in Figure~\ref{fig:burst_slope}, the rise times are rapid, with fades that are slower but still on timescales associated with the inner disk.  Second, the accretion rates of $>10^{-7}$ M$_\odot$ yr$^{-1}$ during large outbursts are near the range where the optical and near-IR spectrum should start to show evidence for accretion heating of the viscous disk, including as CO absorption \citep{liu22}, if the system were in steady state.  The lack of these features indicates that the innermost disk is not being fed at such a large rate, otherwise the size of the heated disk would be larger and would dominate the spectrum.   Once the excess mass in the innermost disk is accreted, the outburst fades.

The EX Lup-type outbursts occur as gas builds up at or very near the inner disk and is then released onto the star. In the framework of  \citet{dangelo10,dangelo12}, the magnetic field of the star connects to the disk at a truncation radius that is larger than the corotation radius, thereby preventing gas from accreting.  Gas piles up at the truncation radius until the truncation radius moves interior to the corotation radius, producing a large flow of gas onto the star.  An alternative hypothesis proposed by \citet{armitage16} suggests that cycles of the stellar dynamo can trigger a burst of accretion through a change in polarity, which generates a more efficient Hall effect.

Our results pose challenges for both of these interpretations.  For the magnetospheric instability model, the accretion should cease until the truncation radius gets smaller, but EX Lup during quiescent periods accretes gas at a similar rate as other accreting young stars.  In this case, the truncation radius must be leaky.  Moreover, whatever mechanism that causes gas to build up at the star would probably be common.  

For the stellar dynamo triggering hypothesis, large outbursts should occur at the (semi-) regular intervals expected for changes in polarity, but outbursts occurred in 1945, 1955, and 2008, not at regular intervals.  In this scenario, the triggering of characteristic and large bursts would have to be distinct, with the characteristic outbursts unexplained.  The accretion burst should begin where the Hall effect is most efficient, beyond the innermost disk.

Whatever the physics, the phenomenon must be a cyclical build-up and release of gas at the inner disk.  During quiescence, the star-disk accretion rate is lower than the accretion rate through the disk.  This process is similar to the toy model that \citet{leeyh20} described for the periodic variable EC 53 (V371 Ser), except that outbursts of EX Lup are much larger and less frequent. 

\subsection{The Phase Lag of Rotational Modulation}
\label{sec:phase-lag-discussion}

The periods measured between 2004 and 2023 for light curves and also from absorption lines from \cite{Kospal2014} remain steady at $\sim 7.415$ days (Table \ref{tab:sin_period}), during periods of weak and strong accretion.  The periodicity is interpreted by \cite{Sicilia-Aguilar2012} and \cite{Sicilia-Aguilar2015} as a stable and rotating, non-axisymmetric accretion column and hot spot tied to the stellar surface. In Section \ref{sec:spot} we described the characteristics of spot modulation in EX Lup multi-color light curves. Here we summarize them as follows

\begin{itemize}
    \item The profile of spot modulation follows sinusoidal waves. The peak-to-peak amplitude of modulation increases to shorter wavelengths, from 0.20 mag ($\sim 10\%$) in the I band to 0.6 mag ($\sim 75\%$ in flux changes) in the U band.   
    \item The light curves exhibit a wavelength-dependent phase-lag, with $I$ band lagged by $\Delta \phi \sim 0.23$ (1.5 days) compared to the $U$ band in post-burst light curves. In-burst light curves have phase lags as well, though smaller than those in post-burst light curves.
\end{itemize}

Following the interpretation of \cite{Sicilia-Aguilar2015}, we develop a toy model with a cool and a hot spot to explain the phase lag and change in phase during 2022.  Although a detailed description of spot parameters is beyond the scope of this study, we present a nominal model that qualitatively reproduces the observed lag in light curves. 
A single circular spot on the surface will produce rotational modulation that is the same at all wavelengths and is therefore insufficient to explain the phase differences. 
With multiple spots, the modulations generated by each spot may have different phases, depending on the longitudes and temperatures.
Two circular cool spots (or two hot spots) would produce entirely anti-phase rotational modulation, when their longitude difference is $180^\circ$, while the longitude difference of one cool and hot spot should be $0^\circ$ to produce the anti-phase modulations. 
The phase lag from multi-color photometry would be the result of color-dependent modulation amplitudes between spots. The spots with different temperatures, thus different spot contrast to the photosphere, would also change the overall modulation phase at different wavelengths.

\begin{figure*}
    \centering
    \includegraphics[width=0.95\textwidth]{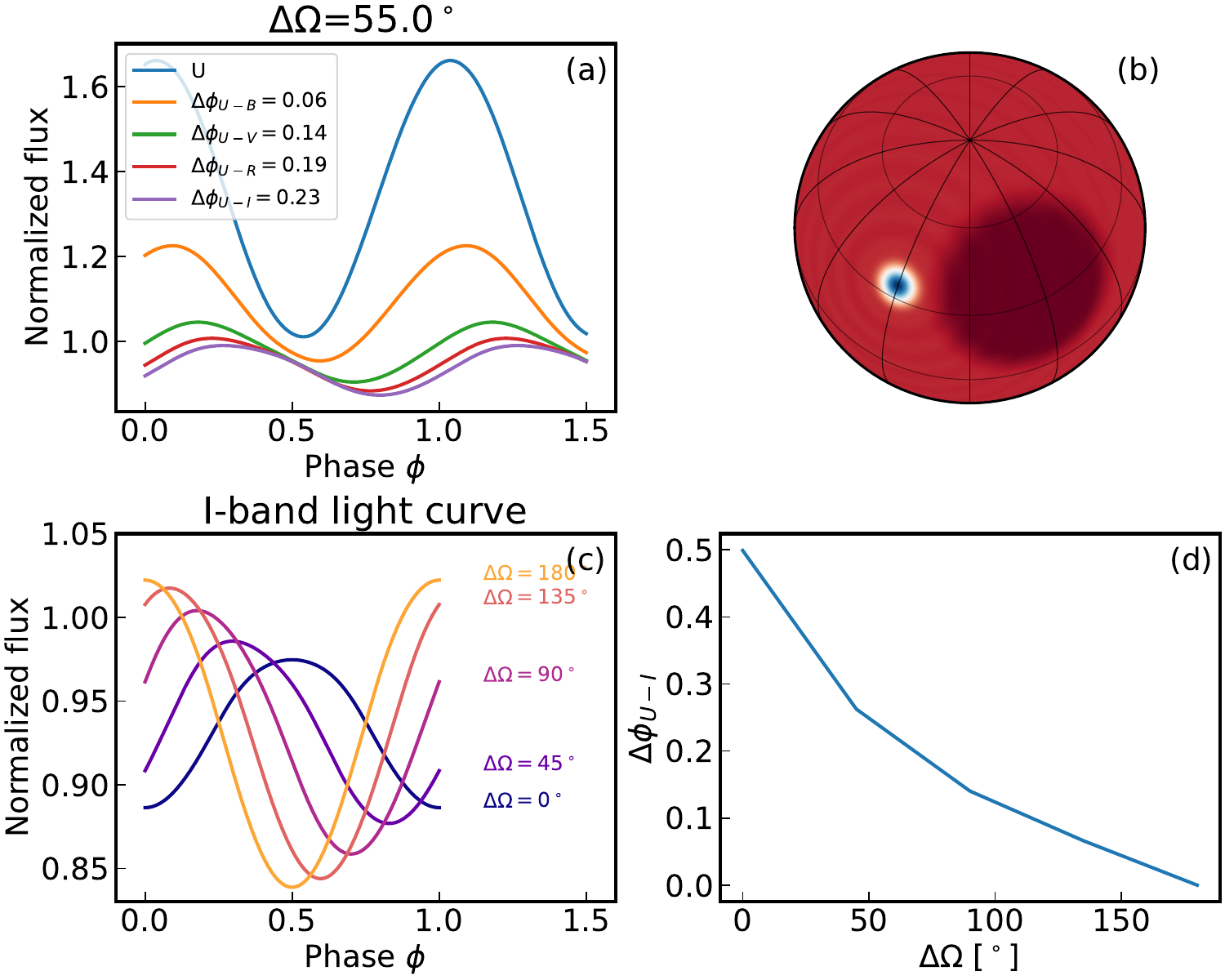}
    \caption{A schematic illustration of a two-component spot model that can produce the observed multi-color light curves. (a) The multi-color light curves from the two-component spot model. The predicted phase lags between U and other bands are indicated. (b) The intensity plot of the stellar surface for the two-component spot model, with hot (blue) and cool (dark red) spots, and uniform photosphere background (red). The adopted model parameters are listed in Table \ref{tab:two-component_spot_model}. (c) The phase shift and morphological change of synthetic I band light curves according to different longitude differences of two spots $\Delta \Omega$.  (d) The resulting phase lags of brightness maxima between U and I band light curves as a function of spots' longitude differences. This set of figures is produced by the python package \texttt{starry} \citep{luger2019}.} 
    \label{fig:spot}
\end{figure*}

A schematic illustration of the two-component spot model and its synthetic light curves are summarized in Figure \ref{fig:spot}. In our proposed model, EX Lup has a rotation period of 7.42 days and an inclination of 30$^\circ$ (as estimated by \citealt{Sipos2009} and \citealt{Sicilia-Aguilar2015} from accretion diagnostics, consistent with the $32.4^\circ$ inclination of the cold dust disk imaged with ALMA by \citealt{hales18}). The photosphere temperature is adopted as $T_{\rm pho}=3850~K$ \citep{alcala17}, with two adjacent hot and cool spots with large temperature differences (Figure \ref{fig:spot} b). For simplicity, we adopt blackbody emission to calculate the spot-photosphere contrast at different wavelengths. However, the $\sim 0.4$ mag amplitude in the TESS light curve indicates that the cool spot must cover $\sim50$\% of the visible stellar surface (see, e.g., 80\% coverage fraction for a similar amplitude variability of LkCa 4, measured by \citealt{gully17}). The hot spot is related to a confined area at the footpoint of the accretion column \cite[see, ][]{Grosso2010,Sicilia-Aguilar2015}, so the size of the hot spot should be smaller than the proposed cool spot, but with much higher temperature compared to the photosphere to induce significant flux variations in short wavelengths.
Due to the color difference and the large size of the cool spot, the amplitudes of the combined spot modulations differ between wavelengths, with the U-band light curves largely modulated by the hot spot and the I-band light curve more influenced by the cool spot (Figure \ref{fig:spot} b). 

The simulated phase-lag between $U$ and $BVRI$ band is similar to the observed ones (Figure \ref{fig:phasefolded_lightcurve}). The model parameters, listed in Table \ref{tab:two-component_spot_model}, serve only as a guide.  The phase lag and amplitudes have a degenerate solution in spot contrast (the spot temperature) and their distribution in longitude, as seen in Figure \ref{fig:spot} c, where we show how the phase and morphology of combined I-band light curves change with the longitude difference between two spots.  

This two-spot model for the periodicity of EX Lup, inferred from the phase lag, is supported by the change in phase from before and after the 2022 burst (see Figure \ref{fig:asassn_phasefolded_lightcurve} in \S \ref{sec: asassn_tess_spot}).  Before the 2022 burst, the accretion was weaker, so the periodicity was modulated by the rotation of the cool spot.  After the 2022 burst, the accretion is stronger, so the rotational modulation is determined by the rotation of the hot spot, at a different phase than the cool spot. 
The stochastic bursts gradually develop in the modulation minima between 2019-2021 (as is also seen in TESS light curve Figure \ref{fig:tess}), so the region hosting these bursts should be close to the large cool spot in the longitude phase, as visualized in Figure \ref{fig:spot}.

A phase lag of $\Delta \phi_{U-I} \sim 0.05$ was measured previously for the active star HR 1099, interpreted as two cool spots and one hot spot on the stellar surface \citep{Zhai1994}.  More relevant analogs are accreting young stars, which feature photospheric cool spots and magnetic hot spots.  \cite{Espaillat2021} measured a time-lag of a tenth of the rotational period between UV and optical peaks for the classical T Tauri star GM Aur.  In a follow-up analysis, \cite{Robinson22} found time lags in TESS and UBVRI light curves for 14 classical T Tauri stars, suggesting these longitudinal stratifications of accretion columns might be common in the CTTSs.  
The time lag is interpreted by the non-azimuthally symmetric accretion hot spot region that has a radial density gradient distribution, leading to different SED profiles at different rotational phases.  Although this description seems discrepant with our interpretation and model, the formulation and interpretation are actually quite similar to the presence of hot and cool spots.

\begin{deluxetable}{lcc}
\tablecaption{Comparisons between EX Lup's Light Curve Modulation Periods and RV period.}
\label{tab:sin_period}
\tablehead{\colhead{Time} & \colhead{Data} & \colhead{Periodogram Period}\\
		   \colhead{} & \colhead{} & \colhead{(day)}}
\startdata
2004-2005				&          $V$                          &	7.406$\pm$0.098   \\
2007-2012				&	Absorption lines \tablenotemark{a}	&	7.417$\pm0$.001	\\
2016-2018				&          $V$                          &	7.419$\pm$0.061   \\
2018-2022				&          $g$                          &	7.412$\pm$0.042   \\
2022-2023				&          $g$                          &	7.411$\pm$0.102   \\
\enddata
\tablenotetext{a}{While the periodicities from emission lines are more scattered between 6-8 days, see the Table A.2 in \cite{Sicilia-Aguilar2015} for details  }
\end{deluxetable}

\begin{deluxetable}{lcccc}
\tablecaption{The two-component spot model of Figure \ref{fig:spot}}
\label{tab:two-component_spot_model}
\tablehead{  & \colhead{$T_{\rm spot}$} & \colhead{Radius} & \colhead{Latitude} & \colhead{Longitude}} 
\startdata
 Hot spot	&	6500 K	&	4$^\circ$	&	35$^\circ$	&	0$^\circ$ \\
 Cool spot	&	3000 K	&	30$^\circ$	&	35$^\circ$	&	55$^\circ$\\
\enddata
\tablecomments{ The effective temperature of photosphere and the inclination of the star is assumed to be 3850 K and 30$^\circ$.}

\end{deluxetable}

\section{Conclusions\label{sec:conclusion}}

In this work, we have assembled the EX Lup light curve from 1895 to 2023 to assess the role of accretion bursts and outbursts in this archetype of its class.  The brightness increases from historical photometric data are divided into two categories: the major outbursts (occurred in 1945, 1955, and 2008) and characteristic bursts (14 detected in the past 70 years). Three outbursts occurred during 130 years of observations, each with large photometric variations up to 5 mags and year-long durations. 

We build the historical record of accretion onto EX Lup in Table \ref{tab:burst_info} by leveraging correlations during multi-band photometry of the characteristic burst in 2022, along with past flux-calibrated spectra.  The accretion rate reaches a peak of $\sim 5\times10^{-7}$ \mdotyr\ during large bursts and $\sim 2-5\times10^{-8}$ \mdotyr\ during characteristic bursts, and wanders from $10^{-9}$ to $10^{-8}$ \mdotyr\ during quiescent periods of steady accretion.
The two outbursts in 1955 and 2008 accrete $\sim$0.1 Earth masses, or
around 80\% of the total burst-consumed masses in seventy years, while the characteristic bursts are more common than the outburst, but they typically only consume a few $10^{-3}$ Earth masses during the burst.  The burst-consumed mass is roughly 2 times the mass accreted during quiescent periods, a comparison limited by uncertainty in the frequency of large bursts and a more precise long-term baseline for quiescent accretion rate.

The characteristic burst in 2022 and the quiescent periods before and after the burst feature spot modulation with a periodicity of $\sim$7.415 days, revealed through consistent, daily monitoring by ASAS-SN and AAVSO.  
These observations are consistent with the scenario that a hot spot originated in the shock region linking the stellar surface and the non-axisymmetric accretion column, which is modulated by the stellar rotation \citep[see, e.g., ][]{Sicilia-Aguilar2015}. Moreover, color-dependent time lags and phase shifts from 2022 post-burst multi-color light curves indicate the presence of both a cool and a hot spot.

Our methodology based on photometry to estimate accretion rates provides an approach to exploit the historical data for long-monitored young variable stars.  New surveys such as LSST, WFST, and TiDO, combined with ASAS-SN and AAVSO observers will 
open a new dimension to compare the characteristics of eruptive young stars \citep[see, e.g.,][]{bonito23}: do other stars that historically show repetitive eruptions, especially EXors, also exhibit two-mode bursts and accrete a significant amount of masses in bursts during their final stages?

As for EX Lup itself, we expect several more characteristic bursts in the next decade, based on the clustering of previous bursts between 1993-2002 and the gradually elevated activities seen from recent light curves.  Since the 1955 burst occurred only 10 years after the 1945 burst, we should trigger additional spectroscopic observations early to follow the accretion and ejection processes during the burst, since the physics and the consequences for the disk are studied but still insufficiently explained.  Even if we have to wait until 2060 for the next major outburst, continued monitoring will allow us to assess why mass builds up in the inner disk and why, eventually, that dam will break.

\section{Acknowledgements}

We thank the anonymous referee whose comments improve the quality of this 
paper. We are extraordinarily grateful for the extensive contributions by many people, most especially Albert Jones, for historical monitoring of EX Lup.
We thank Kenneth Menzies and other observers for contributions to the AAVSO light curve during the 2022 burst.  We acknowledge with thanks the variable star observations from the AAVSO International Database contributed by observers worldwide and used in this research.

We also thank Carlo Manara for discussions and providing the X-Shooter spectrum obtained in 2010 and Nienke van der Marel for contributions to the WHT/ISIS spectrum obtained in 2012.   GJH thanks Sergei Nayakshin and Guoxingan Ma for the discussions about this project.

MW and GJH appreciate the Chinese Center for Advanced Science and Technology for hosting the Protoplanetary Disk and Planet Formation Summer School in 2022, organized by Xue-ning Bai and Ruobing Dong, which facilitated this research.  GJH also appreciates discussions with Will Fischer, Lynne Hillenbrand, Agnes Kospal, and Mike Dunham during the preparation of Protostars and Planets VII review, which helped to motivate this work.

This work is supported by the National Natural Science Foundation of China (grant Nos. 11973028, 11933001, 1803012, 12150009, 12173003) and the National Key R\&D Program of China (2019YFA0706601, 2022YFA1603102). We also acknowledge the Civil Aerospace Technology Research Project (D050105). D.J.\ is supported by NRC Canada and by an NSERC Discovery Grant.

This paper includes data collected with the TESS mission, obtained from the MAST data archive at the Space Telescope Science Institute (STScI). Funding for the TESS mission is provided by the NASA Explorer Program. STScI is operated by the Association of Universities for Research in Astronomy, Inc., under NASA contract NAS 5-26555.  The TESS data presented in this paper were obtained from the Mikulski Archive for Space Telescopes (MAST) at the Space Telescope Science Institute. The specific observations analyzed in Sector 12 and 39 can be accessed via MAST \citep{t9-nmc8-f686}.

\bibliography{new.ms}
\bibliographystyle{aasjournal}

\appendix

\section{Extinction Analysis of EX Lup}

In this appendix, we present our combined photosphere and accretion continuum fits to the VLT/X-Shooter spectrum, with an important downward revision to the measured extinction.  Briefly, we fit the EX Lup X-Shooter spectrum with several photospheric templates to simultaneously determine the best extinction and photosphere spectral type for EX Lup.

The first stage of the fitting is carried out individually to three photospheric templates, TWA 25 (M0), TWA 14 (M0.5), and TWA 13A (M1), obtained from \cite{Manara2013}.  For each template, the initial fit has two free parameters: the extinction of EX Lup $A_V$ and a flux scaling factor $s$ of the photospheric template.
We first de-redden the observed EX Lup spectrum $f_{\rm EX~Lup,~obs}$ with $A_V$, using $R_V=3.1$ reddening law \citep{ccm89}. The photospheric template $f_{\rm pho}\times s$ is then subtracted from the dereddened spectrum $f_{\rm EX~Lup,~unred}$ to yield the accretion continuum, $f_{\rm acc}$, as follows:
\begin{equation}
    f_{\rm acc} = f_{\rm EX~Lup,~unred} - f_{\rm pho}\times s
\end{equation}
If we assume the accretion fluxes should be constant continuum over 4000-6000 \AA, then the 
$\chi^2$ expression is defined as the goodness-of-fit
\begin{equation}
\label{eq:app_goodness_of_fit}
    \chi^2 = \sum_{\lambda}\frac{\big( f_{\rm acc,~4000-6000}-\texttt{med}(f_{\rm acc,~4000-6000}))^2}{(0.2\times f_{\rm EX~ Lup,~obs,~4000-6000})^2},
\end{equation}
with $\texttt{med}(f_{\rm acc,4000-6000})$ as the median value of accretion fluxes between 4000-6000 \AA. The uncertainty in fluxes is treated as the 20\% of fluxes of the observed EX Lup spectrum $f_{\rm obs}$. 
The best-fit model is found by minimizing the $\chi^2$ defined by Eq. \ref{eq:app_goodness_of_fit} in the range of $A_V=[0.0,1.2]$ mag with steps of 0.1 mag. The range of scaling factor $s$ differs depending on the templates being used.

\begin{figure}
    \centering
    \includegraphics[width=.9\linewidth]{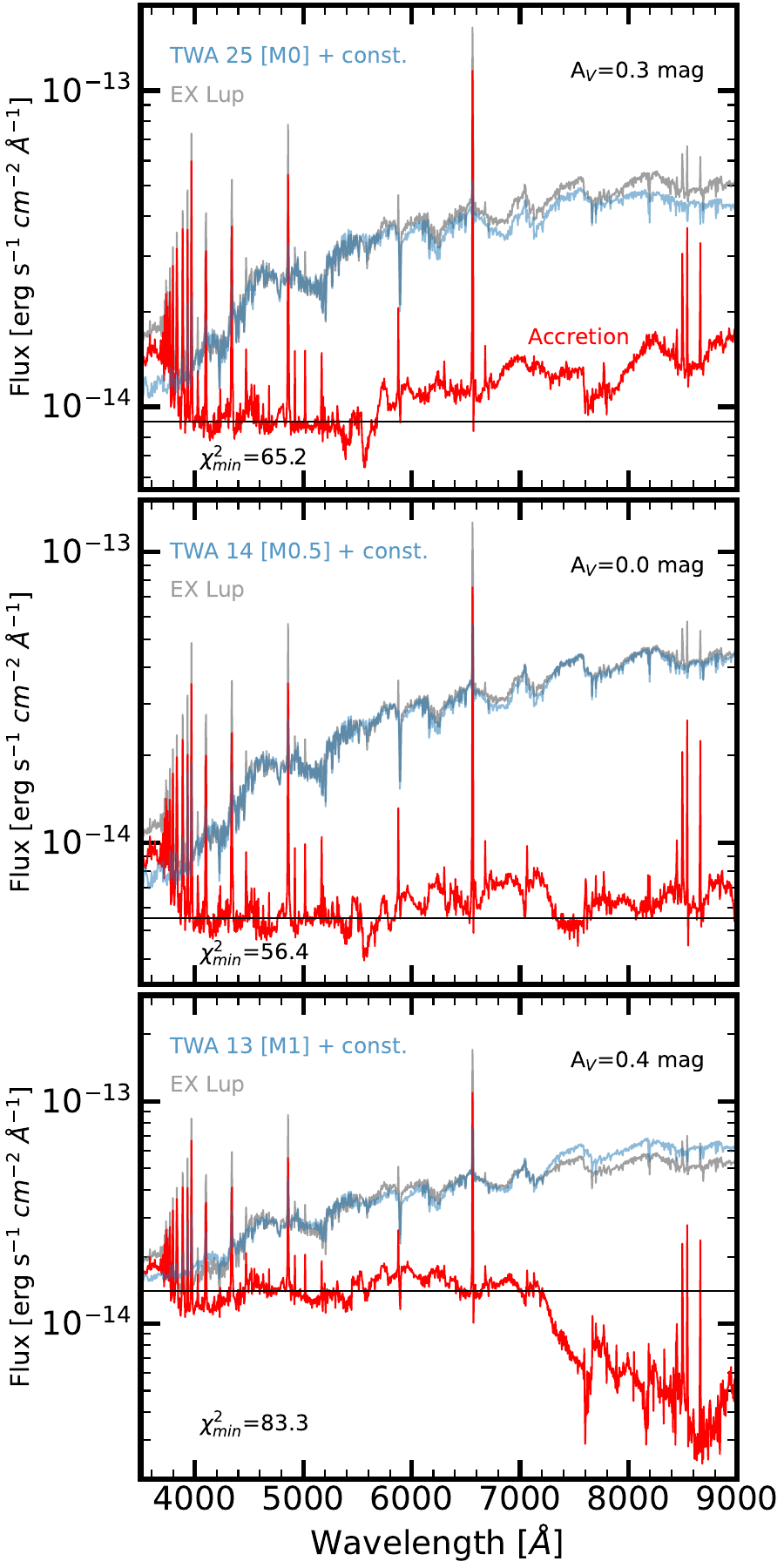}
    \caption{Comparisons of EX Lup VLT/X-Shooter spectrum with best-fit extinction $A_V$ for three photosphere templates. The photosphere templates are taken from the non-accreting YSOs' VLT/X-Shooter spectra of \cite{Manara2013}, with names and spectral types indicated on the upper left. We assume $A_V=0$ for these photosphere templates. The best-fit $A_V$ for each template is obtained by minimizing the standard deviation (listed lower left) of accretion fluxes (red) between 4000-6000 $\AA$ and is also displayed on the upper right. The EX Lup spectrum (gray) is dereddened with best-fit $A_V$ in each panel. The templates are added with constant fluxes (black) to compare with EX Lup spectrum.}
    \label{fig:app_temp_comparison}
\end{figure}

Figure~\ref{fig:app_temp_comparison} compares the best-fit $A_V$ of EX Lup for three photospheric templates. The best-fit $A_V$ of EX Lup is 0.3, 0.0, and 0.4 mag when using TWA 25, TWA 14, and TWA 13 as the photospheric templates, respectively. 
A change of 10\% $\chi^2_{min}$ leads to an uncertainty of $A_v$ of 0.2-0.3 mag and a 20-30\% change of photosphere fluxes. Among the three candidate templates, TWA 14 is the optimal template that yields the lowest $\chi^2$ value. Both the TWA 13 and TWA 25 spectra have a different slope from dereddened EX Lup spectrum at $\lambda>6500~\AA$, resulting in lower and higher accretion fluxes than the accretion continuum at 4000-6000 $\AA$. 

However, TWA 14 is shallower in the depths of several absorption bands between 5500-7500 $\AA$, compared with EX Lup. Since the depth of molecular absorption bands is inversely proportional to the effective temperature of the photosphere, and therefore the spectral type, we infer that EX Lup has a spectral type between TWA 14 and TWA 25.

In the second stage of fitting, we interpolate a suite of synthetic photospheric templates representative of spectral types between M0 and M0.5, using a mixture of TWA 14 and TWA 25 spectra with varying weights and search for a minimum $\chi^2$ on a new [$A_V$, $s$, $w$] grid.

\begin{figure}
    \centering
    \includegraphics[width=0.9\linewidth]{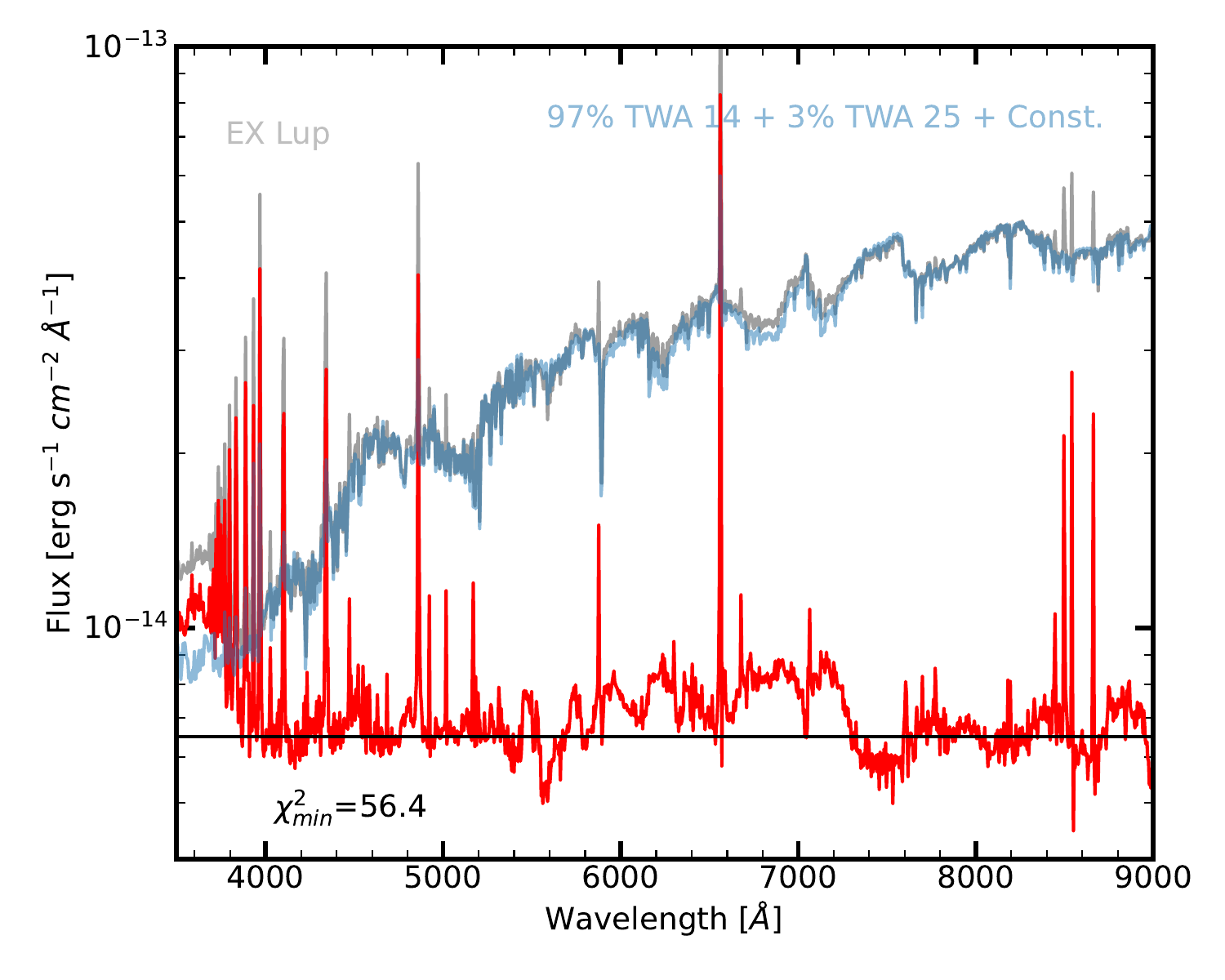}
    \caption{The EX Lup spectrum dereddened with best-fit $A_V$ using a mixture of TWA 14 and TWA 25 spectra. The best-fit $A_V$ is 0.1 mag with synthetic spectrum made up of 97\% TWA 14 and 3\% TWA 25.}
    \label{fig:app_exlup_bestfit}
\end{figure}

The new best fit of $A_V$ is 0.1 mag with a photosphere template consisting of 97\% TWA 14 fluxes and 3\% TWA 25 fluxes. The best model is shown in figure \ref{fig:app_exlup_bestfit}. Among the top-ranking models ($\Delta \chi^2/\chi^2_{min} = 10 \%$), the range of $A_V$ and weights of TWA 14 spans 0.0-0.2 mag and 0.8-1.0, the latter implying M0.4-0.5 spectral type of EX Lup. Models with earlier spectral types favor larger $A_V$. 

The $\chi^2$ of the best-fit model, 56, is not significantly improved compared to the best model using TWA 14 as the template in Figure \ref{fig:app_temp_comparison}. In the new best-fit model, accretion fluxes at $\lambda > 7500 \AA$ are more aligned with the fluxes at 4000-6000 \AA, though the inconsistencies in molecular band depths between 5500-7500 $\AA$ still remain. These inconsistencies might come from causes other than the photospheric spectral types.

\begin{figure}
    \centering
    \includegraphics[width=0.9\linewidth]{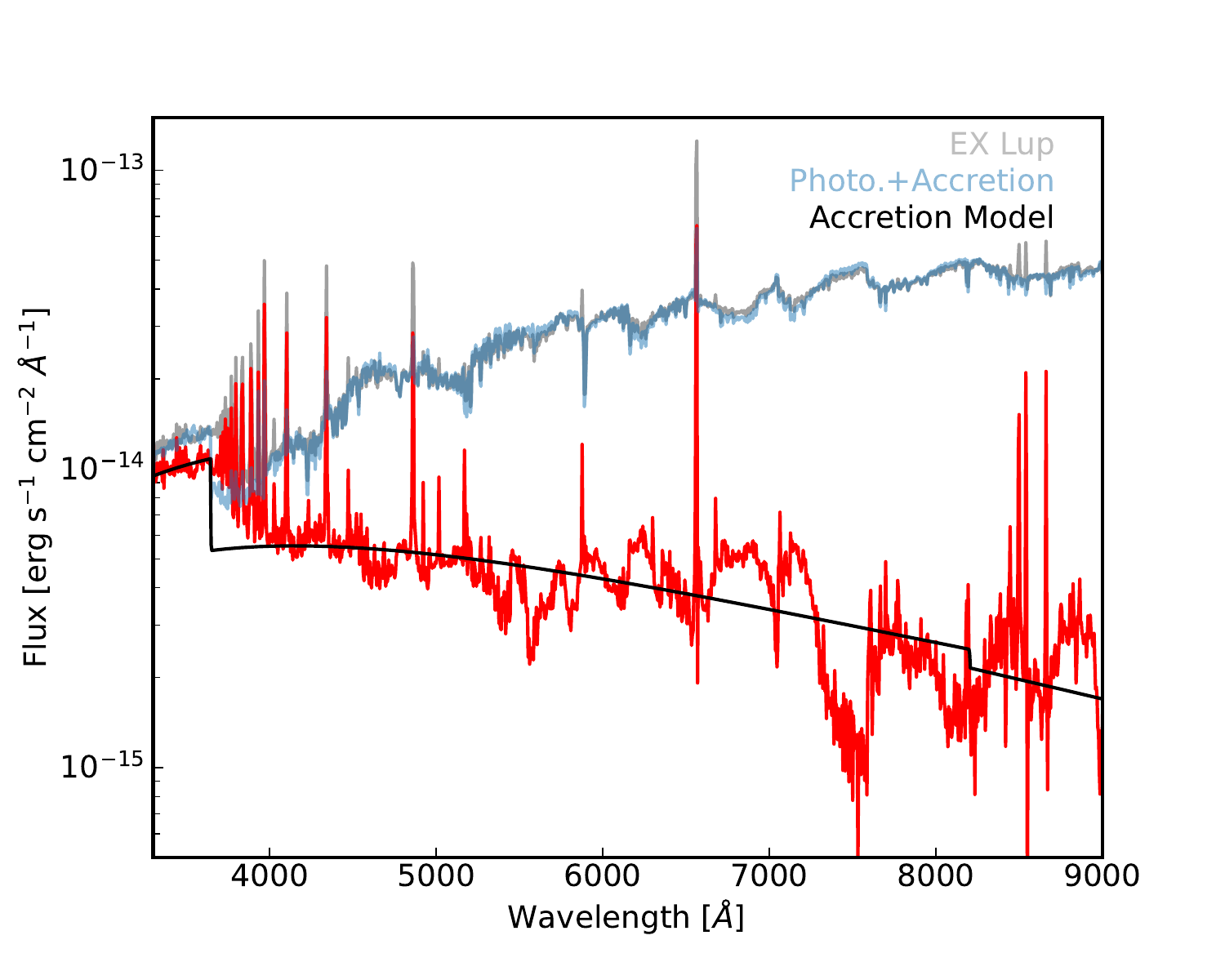}
    \caption{The same as Figure \ref{fig:app_exlup_bestfit} but used an accretion model instead of assuming constant accretion fluxes. The best-fit $A_V$ and spectral type for EX Lup is 0.1 mag and M0.45. The parameters are adopted throughout the analysis of this paper.}
    \label{fig:app_exlup_bestfit_accmodel}
\end{figure}

We also tried a set of fits using plane parallel slabs to estimate the accretion \citep{valenti93}.  These slabs have bluer spectra than the constant spectrum.  We again find a best fit of $A_V$ = 0.1 mag with spectral type M0.45 (Figure \ref{fig:app_exlup_bestfit_accmodel}). The photospheric magnitudes, $A_V$, and accretion spectrum used throughout the paper are taken from this best-fit result.

To summarize the result of our EX Lup extinction analysis, we find that EX Lup spectra can be best fitted with $A_V = 0.1$ mag, using a composite photospheric template with a spectral type of M0.4-0.5. We adopt $A_V = 0.1$ mag and M0.5 throughout this paper.

\section{Additional optical spectra of EX Lup}

We also report several low-resolution optical spectra of EX Lup obtained during quiescence, with flux calibration that was inadequate for full spectral fits.  A WHT/ISIS spectrum covering 3200-10000 \AA\ at $R\sim1000$ was obtained on MJD 56144.869 at an airmass of 2.84, which prevented confidence in our flux calibration.  The spectrum shows a Balmer Jump with an accretion rate of 2--3 $\times10^{-9}$~M$_\odot$ yr$^{-1}$ and with a broadband flux that is consistent with the $\sim M0.5$ spectral type and low extinction.

We also obtained spectra on MJD 57155 and 57195 with the LCO 2m telescope at  Siding Spring Observatory.  Neither the spectra could not be flux calibrated because the observations were separated from the flux calibrator and other problems with the instrumental setup.  However, both show the Balmer Jump, consistent with ongoing accretion, as well as the molecular features consistent with the late-K/early-M spectral type.

\end{document}